\documentclass{jnmp}

\usepackage{amsmath}

\JNMPnumberwithin{equation}{section}

\newtheorem{theorem}{Theorem}[section]
\newtheorem{lemma}{Lemma}[section]

\begin{document}

\renewcommand{\evenhead}{S~N~M Ruijsenaars}
\renewcommand{\oddhead}{Reflectionless Analytic Difference Operators
 III.~Hilbert Space Aspects}

\thispagestyle{empty}

\FirstPageHead{9}{2}{2002}{\pageref{Ruijsenaars-firstpage}--\pageref{Ruijsenaars-lastpage}}{Article}

\copyrightnote{2002}{S~N~M Ruijsenaars}

\Name{Reflectionless Analytic Difference Operators \\
 III.~Hilbert Space Aspects}\label{Ruijsenaars-firstpage}

\Author{S~N~M RUIJSENAARS}

\Address{Centre for Mathematics and
Computer Science,\\
 P.O.Box 94079, 1090 GB Amsterdam, The Netherlands}

\Date{Received November 15, 2001; Accepted March 12, 2002}

\begin{abstract}
\noindent
In the previous two parts of this series of papers, we introduced and
studied a large class of analytic
difference operators admitting reflectionless eigenfunctions, focusing on
algebraic and function-theoretic
features in the first part, and on connections with solitons in the second
one. In this third part we study our
difference operators from a~quantum mechanical viewpoint. We show in
particular that for an arbitrary difference
operator $A$ from a certain subclass, the reflectionless $A$-eigenfunctions
can be used to construct an
unbounded self-adjoint reflectionless operator $\hat{A}$ on $L^2({\mathbb R},dx)$,
whose action on a suitable core
coincides with that of $A$.
\end{abstract}

\section{Introduction}

In this paper we study various quantum mechanical features of a
large class of analytic difference operators that admit
reflectionless eigenfunctions. Our analytic difference ope\-rators
(from now on A$\Delta$Os) are given by
\begin{equation}\label{A}
A\equiv \exp
(-i\partial_x)+V_a(x)\exp(i\partial_x)+V_b(x),
\end{equation}
where $V_a(x)$ and $V_b(x)$ are meromorphic functions with asymptotics
\begin{equation}\label{Vas}
\lim_{|{\rm Re}\, x|\to\infty}V_a(x)=1,\qquad
\lim_{|{\rm Re}\, x|\to\infty}V_b(x)=0.
\end{equation}
The notion `reflectionless eigenfunction' refers to meromorphic functions
${\mathcal W}(x,p)$ satisfying the eigenvalue
equation
\begin{equation}\label{AW}
(A{\mathcal W} )(x,p)=\left(e^p+e^{-p}\right){\mathcal W}(x,p),\qquad p\in{\mathbb C},
\end{equation}
and having asymptotics
\begin{equation}\label{Was}
{\mathcal W} (x,p)\sim \left\{
\begin{array}{ll}
e^{ixp}, & \text{\rm Re}\, x\to\infty,  \vspace{1mm}\\
a(p)e^{ixp},  & {\rm Re}\, x\to -\infty.
\end{array} \right.
\end{equation}

We make extensive use of previous results in this series of
papers, denoting Refs.~\cite{r=0I} and \cite{r=0II} by Part~I and
Part~II, resp. In Part~I we presented and studied a huge class of
reflectionless A$\Delta$Os, but here we are concerned with a far
smaller class. Indeed, our main interest in this paper is in
associating with the A$\Delta$O $A$ a well-defined self-adjoint
operator~$\hat{A}$ on the Hilbert space ${\mathcal H}_x$, where we
use the notation
\begin{equation}\label{Hy}
{\mathcal H}_y\equiv L^2({\mathbb R},dy),
\end{equation}
and we are only able to do so by imposing drastic restrictions on the
spectral data in terms of which the
coefficient functions (`potentials') $V_a$ and $V_b$ are defined.

It should be mentioned at the outset that we are dealing with
exotic territory. In Ref.~\cite{hilb} we studied in great detail
some quite special reflectionless A$\Delta$Os arising in the
context of (reduced, 2-particle) relativistic Calogero-Moser
systems. (See our contribution Ref.~\cite{NE00} to the NEEDS~2000
Proceedings for the unitary similarity transformation connecting
Ref.~\cite{hilb} to the present framework.) From the findings
reported in Ref.~\cite{hilb} it is already clear that a~complete
Hilbert space theory for reflectionless A$\Delta$Os is not going
to amount to a straightforward extension of the well-known results
for reflectionless self-adjoint Schr\"odinger and Jacobi
operators. In Sections~3 and~4 of Part~II we have summarized the
latter results, and we have delineated restrictions on the
spectral data for our A$\Delta$Os such that their reflectionless
eigenfunctions can be tied in with the Schr\"odinger and Jacobi
counterparts.

Roughly speaking, we impose similar restrictions in the present
paper. We shall be quite precise in Section~2, but in this
introduction we try and outline our results with a~minimum of
technical detail, as this might obscure the basically simple plan
of this paper. Before sketching the latter, we add some general
remarks yielding more context. To begin with, since we aim to
associate with $A$ a self-adjoint operator $\hat{A}$ on ${\mathcal
H}_x$, it is natural to restrict $V_a$ and $V_b$ such that $A$ is
at least {\em formally} self-adjoint. Thus $V_b(x)$ should be
real-valued for real $x$, and $V_a(x)\exp(i\partial_x)$ should be
equal to its formal adjoint,
\begin{equation}
[V_a(x)\exp(i\partial_x)]^{*}=\exp(i\partial_x)\overline{V_a(x)}=
\overline{V_a(x-i)}\exp(i\partial_x),\qquad x\in{\mathbb R}.
\end{equation}
Hence we need
\begin{equation}\label{fsa}
V_b^{*}(x)=V_b(x),\qquad V_a^{*}(x)=V_a(x-i),\qquad x\in{\mathbb C},
\end{equation}
where the $*$ denotes the conjugate meromorphic function,
\begin{equation}\label{fconj}
f^{*}(x)\equiv\overline{f(\overline{x})},\qquad x\in{\mathbb C}.
\end{equation}

From now on we restrict attention to potentials satisfying~(\ref{fsa}).
Then a natural strategy would be to try
and find a dense subspace ${\mathcal C}$ in ${\mathcal H}_x$ on which $A$ is well defined and
symmetric. Thus, ${\mathcal C}$ should
consist of square-integrable functions $f(x)$, $x\in{\mathbb R}$, that are restrictions
to ${\mathbb R}$ of functions that have
suitable analyticity properties for $|{\rm Im}\, x|\le 1$, so that there is an
unambiguous meaning for $f(x\pm i)$;
then the function
\begin{equation}
(Af)(x)\equiv f(x-i)+V_a(x)f(x+i)+V_b(x)f(x),
 \qquad x\in{\mathbb R},
\end{equation}
should be square-integrable, and one should have
\begin{equation}
(f,Ag)=(Af,g),\qquad f,g\in{\mathcal C}.
\end{equation}

Assuming such a dense subspace has been isolated, one can try and study the
existence and uniqueness of
self-adjoint extensions. Indeed, the symmetric operator $A$ on ${\mathcal C}$ is
unbounded (due to the shifts), so it
might not have any self-adjoint extensions or a (finite- or
infinite-dimensional) family of self-adjoint
extensions.

In any event, assuming some self-adjoint operator $\hat{A}$ has been
associated with $A$ via this procedure, one
can define its being `reflectionless' solely in terms of time-dependent
Hilbert space scattering theory, as
follows.

First of all, there is a natural `free' dynamics $\exp(-it\hat{A}_0)$ with
which the `interacting' dynamics
$\exp(-it\hat{A})$ can be compared. Indeed, the A$\Delta$O
\begin{equation}\label{A0}
A_0\equiv \exp(-i\partial_x)+\exp(i\partial_x)
\end{equation}
gives rise to an obvious self-adjoint operator $\hat{A}_0$ on ${\mathcal H}_x$,
namely, the transform
\begin{equation}\label{hA0}
\hat{A}_0\equiv{\mathcal F}_0M{\mathcal F}_0^{-1}
\end{equation}
of the self-adjoint multiplication operator
\begin{equation}\label{M}
(Mf)(p)\equiv 2\cosh(p)f(p),\qquad f\in{\mathcal D}(M),
\end{equation}
with maximal domain ${\mathcal D}(M)$ under Fourier transformation
\begin{equation}\label{cF0}
{\mathcal F}_0\, :\, {\mathcal H}_p \to {\mathcal H}_x,\qquad
f(p)\mapsto (2\pi)^{-1/2}\int_{-\infty}^{\infty} dp\,
e^{ixp}f(p).
\end{equation}
(Recall our notation (\ref{Hy}).)

Now assume that the (strong) limits of the operator family
$\exp(it\hat{A})\exp(-it\hat{A}_0)$ for
$t\to\pm\infty$ exist and have equal range. Denoting these isometric wave
operators by $W_{\pm}$, the
corresponding $S$-operator
\begin{equation}
S_x\equiv W_{+}^{*}W_{-}
\end{equation}
is unitary. Since it commutes with the free evolution $\exp(-it\hat{A}_0)$,
its transform
\begin{equation}\label{Sp}
S_p\equiv {\mathcal F}_0^{-1}S_x{\mathcal F}_0
\end{equation}
to ${\mathcal H}_p$ is of the form
\begin{equation}\label{TR}
(S_pf)(p)=T(p)f(p)+R(p)f(-p),\qquad f\in{\mathcal H}_p,
\end{equation}
for certain functions $T(p)$, $R(p)$. Then the dynamics $\hat{A}$
is by definition reflectionless when~$R(p)$ vanishes identically.

Our summary of these notions from time-dependent scattering theory (about
which a wealth of pertinent information
can be found in Ref.~\cite{rs3}) serves a twofold purpose. First, it has
enabled us to sketch a {\em general}
scenario in which the concept of `reflectionless self-adjoint A$\Delta$O'
makes sense and can be studied. Second,
we actually follow a quite different strategy in this paper, but
time-dependent scattering theory does play a
crucial role. Thus we are now better prepared to sketch our {\em special}
setting, and compare it to the above
approach.

The main difference consists in our definition of the self-adjoint operator
$\hat{A}$: It hinges on using the
quite special $A$-eigenfunctions ${\mathcal W}(x,p)$. (Note that in the general
setting just sketched, eigenfunctions of
the A$\Delta$O $A$ need not be and are not mentioned.)
Specifically, the eigenfunction transform
\begin{equation}\label{cF}
{\mathcal F} \, :\, {\mathcal H}_p \to {\mathcal H}_x,
\qquad f(p)\mapsto (2\pi)^{-1/2}\int_{-\infty}^\infty dp\, {\mathcal W}(x,p)f(p)
\end{equation}
plays a decisive role in defining $\hat{A}$.

We have already seen the simplest example of this approach. Indeed, for the
A$\Delta$O $A_0$ (\ref{A0}) we defined
the associated Hilbert space operator $\hat{A}_0$ by using the
$A_0$-eigenfunctions $\exp(ixp)$, $p\in{\mathbb R}$,
cf.~(\ref{cF0}). The unitarity of ${\mathcal F}_0$ is crucial here: Invertibility of
${\mathcal F}_0$ would not be enough
for~(\ref{hA0}) to give rise to a self-adjoint operator $\hat{A}_0$. To compare
with the general strategy, we
mention that a domain ${\mathcal C}_0$ of essential self-adjointness as considered
above is for instance given by
${\mathcal F}_0(C_0^\infty({\mathbb R}))$. The point is, however, that the latter domain cannot be
readily described in terms of the
position space ${\mathcal H}_x$. Moreover, even for $A_0$ there exists an
infinite-dimensional family of distinct domains
of essential self-adjointness yielding distinct reflectionless self-adjoint
operators on ${\mathcal H}_x$.
(This can already be concluded from the special cases studied in
Ref.~\cite{hilb}, cf.~also Ref.~\cite{NE00}. The
present more general case yields a much larger family, as shown at the end
of Section~4.)

To appreciate the latter state of affairs, and, accordingly, the {\em
choice} involved in taking~(\ref{cF}) as a
starting point, a crucial feature of $A$-eigenfunctions should be recalled:
They remain eigenfunctions with the
same eigenvalue after they are multiplied by an arbitrary meromorphic
function with period~$i$. In particular,
this entails that when an A$\Delta$O of the form (\ref{A})--(\ref{Vas}) admits
a reflectionless eigenfunction
${\mathcal W}(x,p)$ satisfying (\ref{AW})--(\ref{Was}), it also admits a
reflectionless eigenfunction $\tilde{{\mathcal W}}(x,p)$
with any other function $\tilde{a}(p)$ in its asymptotics (\ref{Was}).
Indeed, we need only set
\begin{equation}
\tilde{{\mathcal W}}(x,p)\equiv \mu(x,p){\mathcal W}(x,p),
\end{equation}
with
\begin{equation}\label{musp}
\mu(x,p)\equiv \left(e^{2\pi x}+\tilde{a}(p)a(p)^{-1}e^{-2\pi x}\right)/
\left(e^{2\pi
x}+e^{-2\pi x}\right),
\end{equation}
to obtain a new eigenfunction with these features.

Now there is no reason to expect that when the operator (\ref{cF}) is
unitary (or at least isometric) for a
particular choice of ${\mathcal W}(x,p)$, it is still unitary/isometric for
eigenfunctions $\tilde{{\mathcal W}}(x,p)$ as just
described. Indeed, in the case of ${\mathcal F}_0$ it can be proved that multipliers
of the form (\ref{musp}) destroy
unitarity. But as already alluded to, in this case there does exist an
infinite-dimensional family of
$i$-periodic multipliers for which unitarity is preserved.

In our approach, then, the Hilbert space features of the eigenfunction
transform ${\mathcal F}$~(\ref{cF}) are of primary
importance. We are able to establish the relevant functional-analytic
features by using the considerable
amount of explicit algebraic and function-theoretic information gathered in
Parts~I and II. In particular, the
surprising connection to classical $N$-particle relativistic Calogero--Moser
systems established in II~Section~5
is instrumental in obtaining important additional information of the same
character, whose derivation we
have relegated to Appendix~A.

In outline, we solve the pertinent Hilbert space problems as follows. First
of all, we choose the spectral data
in terms of which ${\mathcal W}(x,p)$ is defined such that the
transform~${\mathcal F}$~(\ref{cF}) is a bounded operator on
${\mathcal H}_p$. This is already the case whenever ${\mathcal W}(x,p)$ has no poles for
real~$x$, which is a weak
restriction. This choice also ensures that no nontrivial $C_0^\infty({\mathbb R})$-function is
annihilated by ${\mathcal F}$, cf.~Lemma~\ref{lemma:2.1}.
As a consequence, we are entitled to define an operator $\hat{A}$ on the
subspace
\begin{equation}\label{cP}
{\mathcal P} \equiv {\mathcal F} C_0^\infty({\mathbb R}),
\end{equation}
by setting
\begin{equation}\label{hA}
\hat{A}{\mathcal F} f\equiv {\mathcal F} Mf,\qquad
 f\in C_0^\infty({\mathbb R}),
\end{equation}
where $M$ is defined by (\ref{M}). (The relation to the A$\Delta$O $A$ is also
clarified in Lemma~\ref{lemma:2.1}.)

A far more drastic restriction on the spectral data now ensures that
${\mathcal W}(x,p)$ has no poles for ${\rm Im}\, x \in
[-1,0]$. Note that this striking feature is generically destroyed when
${\mathcal W}(x,p)$ is multiplied by
$i$-periodic multipliers $\mu(x,p)$ with constant limits for $|{\rm Re}\,
x|\to\infty$. (Indeed, by Liouville's theorem
the latter must have poles in a period strip to be nonconstant. On the
other hand, these poles might occur at
the same locations as zeros (counting multiplicities) of ${\mathcal W}(x,p)$ in the
strip ${\rm Im}\, x \in [-1,0]$, in which
case
$\mu(x,p){\mathcal W}(x,p)$ would still be pole-free in this strip.)

Due to the absence of these critical poles, we are able to show
that the operator $\hat{A}$ is {\em symmetric} on ${\mathcal P}$.
This involves considerable work, whereas the next step is quite
easy: An application of Nelson's analytic vector
theorem~\cite{rs2} yields essential self-adjointness of~$\hat{A}$
on~${\mathcal P}$, cf.~Lemma~\ref{lemma:2.2}. Denoting the
self-adjoint extension by the same symbol, we obtain a unitary
one-parameter group $\exp(-it\hat{A})$ on the closure
$\overline{{\mathcal P}}$ of the subspace ${\mathcal P}$. In
general, this is a proper subspace of ${\mathcal H}_x$ (that is,
in general ${\mathcal F}$ is not onto ${\mathcal H}_x$), and we
now extend~$\hat{A}$ {\em provisionally} to a self-adjoint
operator acting in ${\mathcal H}_x$ by putting it equal to an
arbitrarily chosen self-adjoint operator on the orthogonal
complement ${\mathcal P}^{\perp}$. (At this stage we do not yet
know that the latter space is spanned by eigenfunctions of the
A$\Delta$O~$A$, so this provisional extension cannot be avoided.)

Our next goal consists in handling the time-dependent scattering theory of
the interacting dynamics
$\exp(-it\hat{A})$, as compared to the free dynamics  $\exp(-it\hat{A}_0)$.
We do this in Section~3, the most
important result being that the wave operators can be written in terms of
${\mathcal F}$, cf.~Theorem~\ref{theorem:3.2}. From our
explicit formulas it is then clear by inspection that ${\mathcal F}$ is an isometry.
Moreover, they show that the
$S$-matrix~$S_p$~(\ref{Sp}) is the one expected from time-independent
scattering theory. (That is, the
$S$-matrix expected from the asymptotics~(\ref{Was}) of the eigenfunction.)

In Section~4 we complete our analysis by clarifying the state of affairs on
${\mathcal P}^{\perp}$: This space is spanned
by finitely many pairwise orthogonal eigenfunctions ${\mathcal W}(x,r_n)$,
$r_n\in i(0,\pi)$, $n=1,\ldots,N_{+}$, with distinct
real eigenvalues $2\cosh(r_n)$. Thus the definition of $\hat{A}$ can be
completed by requiring that its
action on ${\mathcal P}^{\perp}$ be equal to that of $A$, just as its action
on~${\mathcal P}$. The key to understanding
${\mathcal P}^{\perp}$ is an explicit formula for ${\mathcal F}{\mathcal F}^{*}$, which we obtain
along the same lines as similar
formulas for the special cases we studied in Ref.~\cite{hilb}. (Since we
have no duality properties available in
the present general framework, we cannot proceed in this way to obtain the
isometry formula ${\mathcal F}^{*}{\mathcal F}={\bf
1}$. Instead, we exploit the isometry of wave operators, cf.~Section~3.)

We conclude Section~4 with an appraisal of some special cases. Of
particular interest is the subclass for
which no point spectrum occurs in the spectral resolution of $\hat{A}$
(corresponding to $N_{+}=0$). This
infinite-dimensional family has no analog for reflectionless self-adjoint
Schr\"odinger and Jacobi operators. We
use it to illustrate the ambiguity issue discussed above.

\section{Essential self-adjointness on the domain $\boldsymbol{\mathcal P}$}

We begin by recalling how our class of A$\Delta$Os $A$ (\ref{A}) and the
associated reflectionless eigenfunctions
${\mathcal W}(x,p)$ are obtained from `spectral data' $(r,\mu(x))$. The vector
$r=(r_1,\ldots,r_N)$, with $N\in{\mathbb N}^{*}$,
consists of complex numbers satisfying
\begin{equation}\label{r1}
e^{r_m}\ne e^{\pm r_n},\qquad 1\le m<n\le N,
\end{equation}
and
\begin{equation}\label{r2}
{\rm Im}\,  r_n\in \left\{
\begin{array}{ll}
(0,\pi),  &  n=1,\ldots,N_{+},  \vspace{1mm}\\
(-\pi,0),  & n=N-N_{-}+1,\ldots,N,
\end{array} \right.
\end{equation}
where
\begin{equation}
N_{+},N_{-}\in \{ 0,\ldots,N\},\qquad N_{+}+N_{-}=N.
\end{equation}
 The vector $\mu(x)=(\mu_1(x),\ldots,\mu_N(x))$ consists of meromorphic
functions
satisfying
\begin{equation}\label{mu}
\mu_n(x+i)=\mu_n(x),\qquad \lim\limits_{|{\rm Re}\, x|\to \infty} \mu_n(x)=c_n,\qquad
c_n\in{\mathbb C}^*,\qquad
n=1,\ldots,N.
\end{equation}
These `minimal' restrictions on $(r,\mu)$ are in force throughout this
paper. (When the need arises, we specify
additional restrictions.)

Now we define a Cauchy matrix
\begin{equation}\label{C}
C_{mn}\equiv \frac{1}{e^{r_m}-e^{-r_n}},\qquad m,n=1,\ldots,N,
\end{equation}
and a diagonal matrix
\begin{equation}\label{defD}
D(x)\equiv \mbox{diag}\, (d(r_1,\mu_1;x),\ldots,
d(r_N,\mu_N;x)),
\end{equation}
where
\begin{equation}\label{defd}
d(\rho,\nu;x)\equiv \left\{
\begin{array}{ll}
\nu(x)e^{-2i\rho x},  &  {\rm Im}\,  \rho \in(0,\pi),  \vspace{1mm}\\
\nu(x)e^{-2i(\rho +i\pi)x},  & {\rm Im}\,  \rho\in(-\pi,0).
\end{array} \right.
\end{equation}
Then the potentials $V_a$, $V_b$ and wave function ${\mathcal W}$ are defined via the
solution to the system
\begin{equation}\label{sysN}
(D(x)+C)R(x)=\zeta,\qquad \zeta\equiv (1,\ldots,1)^t,
\end{equation}
by
\begin{gather}\label{Va}
V_a(x)\equiv \left( 1+\sum_{n=1}^N e^{r_n}R_n(x)\right) \left(
1+\sum_{n=1}^N e^{r_n}R_n(x+i)\right)^{-1} ,
\\
\label{Vb}
V_b(x)\equiv \sum_{n=1}^N (R_n(x-i)-R_n(x)),
\\
\label{W}
{\mathcal W}(x,p)\equiv e^{ixp}\left( 1-\sum_{n=1}^N
\frac{R_n(x)}{e^p-e^{-r_n}}\right).
\end{gather}

Of course, it is far from obvious that these definitions entail the
eigenvalue equation~(\ref{AW}), but this
is shown in I~Theorem~2.3. In contrast, the asymptotics
\begin{equation}\label{Ras}
\lim_{{\rm Re}\,  x\to\infty}R(x)=0,\qquad \lim_{{\rm Re}\,  x\to -\infty}R(x)=C^{-1}\zeta ,
\end{equation}
easily follows from (\ref{C})--(\ref{sysN}), cf.~also I~Lemma~2.1. Using
(\ref{Ras}), one obtains the asymptotics
(\ref{Vas}) and (\ref{Was}), with
\begin{equation}\label{af}
a(p)\equiv \prod_{n=1}^N\frac{e^p-e^{r_n}}{e^p-e^{-r_n}},
\end{equation}
cf.~I~Theorem~2.3.

We are not able to associate a self-adjoint operator on ${\mathcal H}_x$ to the
A$\Delta$O $A$ unless we impose further
restrictions on the data $(r,\mu)$. But to prove boundedness of the
eigenfunction transform ${\mathcal F}$ (\ref{cF}) and
a few more salient features, we only need a quite weak assumption, as
detailed in the next lemma. (Indeed, for
generic spectral data satisfying (\ref{r1})--(\ref{mu}) the meromorphic
function $R(x)$ has no poles on the
real axis.)

\begin{lemma}\label{lemma:2.1}
Assume that the solution $R(x)$ to (\ref{sysN})  has no poles for real $x$.
Then the operator ${\mathcal F}$ (\ref{cF})
is bounded. For all $\phi\in C_0^\infty({\mathbb R})$
the function $({\mathcal F}\phi)(x)$ extends to a~meromorphic function satisfying
\begin{equation}\label{AM}
A({\mathcal F}\phi)(x)=({\mathcal F} M\phi )(x),\qquad  x\in{\mathbb C},
\end{equation}
where $A$ is the A$\Delta$O (\ref{A}) and $M$ the multiplication
operator~(\ref{M}). Moreover, we have
\begin{equation}\label{KerF}
{\rm Ker}\,({\mathcal F})\cap C_0^\infty({\mathbb R}) =\{ 0\}.
\end{equation}
\end{lemma}

\begin{proof} In view of the asymptotics (\ref{Ras}) and absence of real
poles, the function
$R(x)$ is bounded for real $x$. Due to the restrictions (\ref{r2}), the
functions
$(e^p-e^{-r_n})^{-1}$, $n=1,\ldots,N$, are bounded for real $p$. Hence
${\mathcal W}(x,p)$ (\ref{W}) is bounded for real
$x$,~$p$. Choosing $\phi(p)\in C_0^\infty({\mathbb R})$, we may write
\begin{equation}\label{Frep}
({\mathcal F}\phi)(x)=({\mathcal F}_0\phi)(x) -\sum_{n=1}^N R_n(x)  ({\mathcal F}_0\phi_n)(x),
\end{equation}
with
\begin{equation}\label{phin}
\phi_n(p)\equiv \left(e^p-e^{-r_n}\right)^{-1}\phi(p),
\end{equation}
cf.~(\ref{cF0}). Since Fourier transformation ${\mathcal F}_0$ is a bounded
operator, and the multiplication operators
occurring here are bounded, too, boundedness of ${\mathcal F}$ follows.

To prove the second assertion, we recall the easily verified fact that the
Fourier transform of a
$C_0^\infty({\mathbb R})$-function extends to an entire function. Since we have
$\phi,\phi_1,\ldots,\phi_N\in C_0^\infty({\mathbb R})$, and since
$R_1(x),\ldots,R_N(x)$ are meromorphic, it is clear from (\ref{Frep}) that
$({\mathcal F}\phi)(x)$ extends to a
meromorphic function. The action of~$A$ on this function yields the
meromorphic function
\begin{gather}
({\mathcal F}_0\phi)(x-i)-\sum_{n=1}^NR_n(x-i)({\mathcal F}_0\phi_n)(x-i)
\nonumber\\
\qquad {}+V_a(x)\left[({\mathcal F}_0\phi)(x+i)-
\sum_{n=1}^NR_n(x+i)({\mathcal F}_0\phi_n)(x+i)\right]
\nonumber\\
\qquad {}+V_b(x)\left[({\mathcal F}_0\phi)(x)-\sum_{n=1}^NR_n(x)({\mathcal F}_0\phi)(x)\right].
\end{gather}
For all $x$ for which the functions $R(x)$, $R(x\pm i)$, $V_a(x)$ and $V_b(x)$
have no poles, this
can be rewritten as the absolutely convergent integral
\begin{equation}
(2\pi)^{-1/2}\int_{-\infty}^\infty
 dp[{\mathcal W} (x-i,p)+V_a(x){\mathcal W}(x+i,p)+V_b(x){\mathcal W}(x,p)]\phi(p).
\end{equation}
Thanks to the eigenvalue equation (\ref{AW}), the function in square
brackets amounts to $2\cosh(p){\mathcal W}(x,p)$,
yielding (\ref{AM}).

{\samepage
To prove (\ref{KerF}), we assume $\phi\in C_0^\infty({\mathbb R})$
satisfies ${\mathcal F}\phi =0$. By
(\ref{Frep}) we then have
\begin{equation}\label{Fid}
({\mathcal F}_0\phi)(x)  =\sum_{n=1}^N R_n(x)({\mathcal F}_0\phi_n)(x).
\end{equation}
Consider now the function $e^{ax}R_n(x)$, $x\in{\mathbb R}$,
$a\ge 0$. Due to (\ref{Ras}),
it is bounded for $x\to -\infty$.
The ${\rm Re}\, x\to \infty$ asymptotics of $R(x)$ can be sharpened to an
exponential decay, so that $e^{ax}R(x)$ is
also bounded at $\infty$, provided $a\in [0,c]$, with $c$ small enough.
(The pertinent asymptotic decay easily
follows from (\ref{sysN}), cf.~I(2.41)--(2.42).) Now $({\mathcal F}_0\phi_n)(x)$ is
a Schwartz space function, so it
readily follows that the functions
\begin{equation}
f_n(z)\equiv (2\pi)^{-1/2}\int_{-\infty}^\infty dx e^{-ixz}R_n(x)({\mathcal F}_0\phi_n)(x),\qquad
n=1,\ldots,N,
\end{equation}
are well defined for ${\rm Im}\,  z\in [0,c]$ and analytic for ${\rm Im}\,  z \in (0,c)$.
Moreover, a dominated convergence
argument yields
\begin{equation}
\lim_{a\downarrow 0}f_n(p+ia)=f_n(p),\qquad n=1,\ldots,N,
\end{equation}
uniformly for real $p$.}

From (\ref{Fid}) we deduce that the function
\begin{equation}
f(z)\equiv (2\pi)^{-1/2}\int_{-\infty}^\infty dx  e^{-ixz}({\mathcal F}_0\phi)(x)=\sum_{n=1}^Nf_n(z)
\end{equation}
is also well defined for ${\rm Im}\,  z\in [0,c]$ and analytic for ${\rm Im}\,  z\in(0,c)$.
Since it satisfies
\begin{equation}
\lim_{a\downarrow 0}f(p+ia)=\phi(p),
\end{equation}
uniformly for $p\in{\mathbb R}$, and $\phi(p)\in C_0^\infty({\mathbb R})$ vanishes on an open set,
Painlev\'e's lemma entails $\phi =0$. Hence~(\ref{KerF}) follows.
\end{proof}

In view of (\ref{KerF}), any vector in the subspace ${\mathcal P}$ (\ref{cP}) can be
written as ${\mathcal F}\phi$, $\phi\in C_0^\infty({\mathbb R})$, in
a unique way. Therefore, the operator $\hat{A}$ (\ref{hA}) is a
well-defined linear operator on ${\mathcal P}$, whose
action coincides with that of the A$\Delta$O $A$, cf.~(\ref{AM}). Obviously,
$\hat{A}$ leaves~${\mathcal P}$ invariant.
Moreover, we clearly have
\begin{equation}\label{Anb}
\| \hat{A}^n{\mathcal F}\phi\|\le c^n\|{\mathcal F}\|\,\|\phi\|,\qquad n\in{\mathbb N},
\end{equation}
with $c>0$ depending on supp($\phi$). Thus ${\mathcal P}$ consists of analytic
vectors for $\hat{A}$. In view of Nelson's
analytic vector theorem~\cite{rs2} it now suffices for essential
self-adjointness of $\hat{A}$ on ${\mathcal P}$ that
$\hat{A}$ is symmetric on ${\mathcal P}$.

Next, we detail assumptions on the spectral data that suffice to prove this
critical symmetry property. We do
this in four steps, each of which adds a restriction. This enables us to
use the less restrictive intermediate
assumptions whenever we can show their sufficiency for the result at hand.
(In most cases, however, we do not know
to what extent these assumptions are necessary.)

Our first step consists in imposing {\em formal} self-adjointness~(\ref{fsa}).
This property is ensured by requiring that $r_1,\ldots,r_N$ be purely
imaginary and that the functions
$ie^{-r_n}\mu_n(x)$ be real-valued for $n=1,\ldots,N$ and real $x$,
cf.~I~Theorem~D.1. Our second
step  consists in requiring that  $\mu(x)$ be constant. Thus our second
assumption comes down to
\begin{equation}\label{lan}
ir_n\in{\mathbb R},\qquad ie^{-r_n}\mu_n(x)\equiv \nu_n^{-1}\in{\mathbb R}^{*},\qquad
n=1,\ldots,N.
\end{equation}

Our third requirement reads
\begin{equation}\label{lapo}
\nu_1,\ldots,\nu_N\in(0,\infty).
\end{equation}
There are explicit examples available where
the second assumption (\ref{lan}) and the assumption of Lemma~\ref{lemma:2.1} are
satisfied, but the third assumption~(\ref{lapo}) and symmetry of~$\hat{A}$
are violated. These examples can be gleaned from
Ref.~\cite{hilb}, but we do not spell out the details here.

Our fourth and final restriction can be most easily phrased in terms of the
$\tau$-function
\begin{equation}\label{tau}
\tau(x)\equiv \left|{\bf 1}_N+CD(x)^{-1}\right|.
\end{equation}
In view of our second restriction (\ref{lan}), $\tau(x)$ is an entire
function, cf.~(\ref{defD})--(\ref{defd}).
Likewise, from (\ref{sysN}) we see that $R(x)$ can be written
\begin{equation}\label{RE}
R(x)=E(x)/\tau(x),
\end{equation}
where $E(x)$ is an entire function. Our fourth requirement is now that
$\tau(x)$ have no zeros for ${\rm Im}\,  x\in
[-1,0]$. This entails in particular that $R(x)$ has no real poles (the
assumption of Lemma~\ref{lemma:2.1}).

In Appendix~A we prove that when $N_{+}$ or $N_{-}$ equals $N$ (so that all
$r_n$ lie either on the positive or
on the negative imaginary axis), the third restriction entails the fourth
one, cf.~Lemma~\ref{lemma:A.1}. For $N_{+}N_{-}>0$
this is presumably still true for generic spectral data, but we were unable
to prove this. Explicit
examples we do not present here show that our fourth requirement {\em is}
stronger than the third one. In any
event, in Appendix~A we also prove that for $N_{+}N_{-}>0$ the third
restriction~(\ref{lapo}) together
with the requirement
\begin{equation}\label{alhal}
-ir_1,\ldots,-ir_{N_{+}},-ir_{N_{+}+1}+\pi,\ldots,-ir_N+\pi \in(0,\pi/2]
\end{equation}
are sufficient to obtain the fourth one, cf.~Lemma~\ref{lemma:A.2}.

We continue by showing that our fourth assumption suffices for essential
self-adjointness of~$\hat{A}$ on its
definition domain~${\mathcal P}$. In the proof we use one property of $\tau(x)$ that
cannot be found in Parts~I and II,
namely,
\begin{equation}\label{tst}
\tau^{*}(x)=\tau(x-i).
\end{equation}
This formula is an easy consequence of (\ref{tau}) and the relations
\begin{equation}
\overline{C}=-C^t,\qquad D^{*}(x)=-D(x-i),
\end{equation}
which follow from (\ref{lan}). (Cf.~also Appendix~A and I~Appendix~D.)

\begin{lemma}\label{lemma:2.2}
Assume that the data satisfy (\ref{lan})--(\ref{lapo}) and that $\tau(x)$
(\ref{tau}) has no zeros for ${\rm Im}\,  x\in
[-1,0]$. Then the operator $\hat{A}$ defined by (\ref{hA}) is essentially
self-adjoint on~${\mathcal P}$~(\ref{cP}).
\end{lemma}

\begin{proof} As already detailed, it suffices to prove symmetry of
$\hat{A}$ on ${\mathcal P}$. (Recall the paragraph
containing (\ref{Anb}).)
 For this purpose we fix $\phi_1,\phi_2\in C_0^\infty({\mathbb R})$ and
consider
$({\mathcal F}\phi_1,\hat{A}{\mathcal F}\phi_2)$. By definition, this equals
\begin{gather}
({\mathcal F}\phi_1,{\mathcal F} M\phi_2)    =  \frac{1}{2\pi}
\lim_{R\to\infty}\int_{-R}^R dx\left( \int_{-\infty}^\infty dq
{\mathcal W}(x,q)\phi_1(q)\right)^{-} \nonumber\\
\qquad{}\times \left( \int_{-\infty}^\infty dp
{\mathcal W}(x,p)2\cosh(p)\phi_2(p)\right)
\nonumber\\
\qquad  =
 \frac{1}{2\pi} \lim_{R\to\infty}\int_{-\infty}^\infty dq\overline{\phi_1(q)}\int_{-\infty}^\infty
dp\phi_2(p)\int_{-R}^R dx
\overline{{\mathcal W}(x,q)}2\cosh(p){\mathcal W}(x,p),
\end{gather}
where we used Fubini's theorem in the last step. Rewriting
$(\hat{A}{\mathcal F}\phi_1,{\mathcal F}\phi_2)$ in the same way, we
obtain
\begin{equation}\label{diff}
({\mathcal F}\phi_1,\hat{A}{\mathcal F}\phi_2)
-(\hat{A}{\mathcal F}\phi_1,{\mathcal F}\phi_2)=\frac{1}{2\pi}\lim_{R\to\infty}
\int_{-\infty}^\infty dq\overline{\phi_1(q)}\int_{-\infty}^\infty dp\phi_2(p)
I_R(p,q),
\end{equation}
where
\begin{equation}
I_R(p,q)\equiv \int_{-R}^R dx
\overline{{\mathcal W}(x,q)}{\mathcal W}(x,p)(2\cosh(p)-2\cosh(q)),\qquad p,q\in{\mathbb R}.
\end{equation}

Next, we invoke the eigenvalue equation (\ref{AW}) and the notation
(\ref{fconj}) to rewrite $I_R(p,q)$ as
\begin{gather}
  \int_{-R}^R dx
({\mathcal W}^{*}(x,q)[{\mathcal W}(x-i,p)+V_a(x){\mathcal W}(x+i,p)+V_b(x)
{\mathcal W}(x,p)]
\nonumber\\
 \label{In}
\qquad{}-[{\mathcal W}^{*}(x+i,q)+V_a^{*}(x){\mathcal W}^{*}(x-i,q)+V_b^{*}(x){\mathcal W}^{*}(x,q)]{\mathcal W}(x,p)).
\end{gather}
Recalling (\ref{fsa}), we obtain
\begin{equation}
I_R(p,q)=\int_{-R}^R dx(J(x,p,q)-J(x+i,p,q)),
\end{equation}
where
\begin{equation}\label{J}
J(x,p,q)\equiv {\mathcal W}^{*}(x,q)
{\mathcal W}(x-i,p)-V_a(x-i){\mathcal W}^{*}(x-i,q){\mathcal W}(x,p).
\end{equation}
Now from (\ref{W}) and (\ref{RE}) we have
\begin{equation}\label{Wn}
{\mathcal W}(x,p)=e^{ixp}\left(
1-\sum_{n=1}^N\frac{E_n(x)}{\tau(x)}\frac{1}{e^p-e^{-r_n}}\right),
\end{equation}
with $E_n(x)$ entire. Using (\ref{tst}), we deduce
\begin{equation}
{\mathcal W}^{*}(x,q)=e^{-ixq}
\left(
1-\sum_{n=1}^N\frac{E_n^{*}(x)}{\tau(x-i)}\frac{1}{e^q-e^{r_n}}\right),\qquad
q\in{\mathbb R}.
\end{equation}

From II(2.34) we also have the identity
\begin{equation}\label{Vn}
V_a(x)=\frac{\tau(x+i)\tau(x-i)}{\tau(x)^2}.
\end{equation}

When we substitute (\ref{Wn})--(\ref{Vn}) in $J(x,p,q)$, we can write the
result as
\begin{gather}
J(x,p,q)    =   e^{-ixq}\!\left(
1-\sum_{n=1}^N\frac{E_n^{*}(x)}{\tau(x-i)}\frac{1}{e^q-e^{r_n}}\right)
e^{i(x-i)p}\!\left(
1-\sum_{k=1}^N\frac{E_k(x-i)}{\tau(x-i)}\frac{1}{e^p-e^{-r_k}}\right)\!
\nonumber\\
   \qquad {}-\frac{1}{\tau(x-i)^2}e^{-i(x-i)q}\left(
\tau(x-2i)-\sum_{n=1}^N E_n^{*}(x-i)
\frac{1}{e^q-e^{r_n}}\right)
\nonumber\\
  \label{Jxpq}  \qquad{}\times e^{ixp}\left(
\tau(x)-\sum_{k=1}^NE_k(x)\frac{1}{e^p-e^{-r_k}}\right).
\end{gather}
The point of doing so is that this representation shows that $J(x,p,q)$ has
no poles on and inside the
rectangular contour $\Gamma$ in the $x$-plane connecting $-R$, $R$, $R+i$, $-R+i$.
(Indeed, by assumption $\tau(x)$ is
zero-free for ${\rm Im}\,  x\in [-1,0]$.)

As a consequence, the contour integral
\begin{equation}
C_R(p,q)\equiv \oint_{\Gamma}dxJ(x,p,q)
\end{equation}
vanishes by Cauchy's theorem. On the other hand, $C_R(p,q)$ equals
$I_R(p,q)$ plus the integrals over the
vertical sides of~$\Gamma$. Thus we infer
\begin{gather}
I_R(p,q)=\left( \int_R^{R+i}+\int_{-R+i}^{-R}\right) dx\nonumber\\
\label{Ib}
\phantom{I_R(p,q)=} {}\times
[-{\mathcal W}^{*}(x,q){\mathcal W}(x-i,p)+V_a(x-i){\mathcal W}^{*}(x-i,q){\mathcal W}(x,p)].
\end{gather}
In order to handle the right boundary term for $R\to\infty$, we substitute
\begin{gather}
{\mathcal W}(x,p)\equiv e^{ixp}+\rho_{+}(x,p),\qquad {\mathcal W}^{*}(x,q)\equiv
e^{-ixq}+\rho^{*}_{+}(x,q),\\
\label{Var}
V_a(x)\equiv 1+\rho(x),
\end{gather}
in the first integral of (\ref{Ib}). An inspection of (the proofs of)
Lemmas~\ref{lemma:2.1},~\ref{lemma:2.2} and Theorem~2.3 in Part~I
now reveals
\begin{equation}
\rho_{+}(x,p),\rho_{+}^{*}(x,q),\rho(x)\to 0,\qquad {\rm Re}\, x\to\infty,
\end{equation}
uniformly for $p$, $q$ and ${\rm Im}\,  x$ in ${\mathbb R}$-compacts. An easy dominated
convergence argument then shows that the
contribution to (\ref{diff}) of terms containing at least one~$\rho$
vanishes. Thus we are left with
\begin{gather}
\int_{R}^{R+i}\left(-e^{-ixq+i(x-i)p}+e^{-i(x-i)q+ixp}\right)\nonumber\\
\qquad {}=-4ie^{iR(p-q)}\sinh((p+q)/2)\frac{\sinh((p-q)/2)}{p-q}.
\end{gather}
Substituting this in (\ref{diff}), we can transform to sum and difference
variables to infer that the
contribution of this term vanishes by virtue of the Riemann--Lebesgue lemma.

Next, we substitute
\begin{equation}\label{Wl}
{\mathcal W}(x,p)=a(p)e^{ixp}+\rho_{-}(x,p),\qquad
{\mathcal W}^{*}(x,q)=a^{*}(q)e^{-ixq}+\rho_{-}^{*}(x,q),
\end{equation}
and (\ref{Var}) in the second integral of (\ref{Ib}). Using the
alternative representation I(2.49) of~$a(p)$, we obtain
\begin{gather}
\rho_{-}(x,p)    =   \sum_{n=1}^N\left(\left(C^{-1}\zeta\right)_n-R_n(x)\right)/\left(e^p-e^{-r_n}\right)
\nonumber\\
  \label{rhom} \phantom{\rho_{-}(x,p)  } {} =    \sum_{n=1}^N \left(C^{-1}D(x)R(x)\right)_n/\left(e^p-e^{-r_n}\right),
\end{gather}
where we used (\ref{sysN}) in the second step. As before, we now infer
\begin{equation}
\rho_{-}(x,p),\rho_{-}^{*}(x,q),\rho(x)\to 0,\qquad {\rm Re}\, x\to -\infty,
\end{equation}
uniformly for $p$, $q$ and ${\rm Im}\,  x$ in ${\mathbb R}$-compacts. Thus it remains to
consider the contribution of
\begin{equation}
4ia^{*}(q)a(p)\exp[-iR(p-q)]\sinh((p+q)/2)\sinh((p-q)/2)/(p-q)
\end{equation}
to (\ref{diff}). As before, this vanishes by the Riemann--Lebesgue
lemma. Hence the rhs of~(\ref{diff}) vanishes, entailing symmetry
of $\hat{A}$ on ${\mathcal P}$.
\end{proof}

With the assumptions of the lemma in effect, we can take the closure of the
operator $\hat{A}$ on ${\mathcal P}$ to
obtain a self-adjoint operator. The latter acts on a dense subspace of the
Hilbert space
\begin{equation}
{\mathcal H}_x({\mathcal F})\equiv \overline{{\rm Ran}({\mathcal F})}.
\end{equation}
It follows from the isometry of ${\mathcal F}$, which we prove in the next section,
that the range of~${\mathcal F}$ is actually
closed. At this stage, however, we only know ${\mathcal F}$ is bounded and we have
no information about Ran(${\mathcal F}$) and
its orthogonal complement.

Until further notice, we denote by $\hat{A}$ the self-adjoint operator on
${\mathcal H}_x$ that acts as the closure of
$\hat{A}$ on ${\mathcal H}_x({\mathcal F})$, and as an arbitrarily chosen self-adjoint
operator on the orthogonal complement
${\mathcal H}_x({\mathcal F})^{\perp}$. The results of the next section are independent of
the latter choice.

\section{Time-dependent scattering theory}

Throughout this section, the assumptions of Lemma~\ref{lemma:2.2} are in effect. With
the self-adjoint operator $\hat{A}$ on
${\mathcal H}_x$ defined at the end of the previous section, we show that the wave
operators for the pair of dynamics
$\exp(-it\hat{A})$ and $\exp(-it\hat{A}_0)$ exist and are intimately
related to the eigenfunction transform~${\mathcal F}$.
As a corollary, this yields isometry of ${\mathcal F}$. The following lemma
is the key to these results.

\begin{lemma}\label{lemma:3.1}
For all $\phi\in C_0^{\infty}((0,\infty))$ we have
\begin{gather}\label{pr}
\lim_{t\to \infty}\| ({\mathcal F} - {\mathcal F}_0)\exp(-itM)\phi \| =0,\\
\label{nr}
\lim_{t\to -\infty}\| ({\mathcal F} - {\mathcal F}_0 a(\cdot))\exp(-itM)\phi \| =0,
\end{gather}
where $M$ is given by (\ref{M}) and
$a(\cdot)$ is the operator of multiplication by $a(p)$ (\ref{af}). For all
$\phi\in
C_0^{\infty}((-\infty,0))$ we have
\begin{gather}\label{pl}
\lim_{t\to \infty}\| ({\mathcal F} - {\mathcal F}_0 a(\cdot))\exp(-itM)\phi \| =0,\\
\label{nl}
\lim_{t\to -\infty}\| ({\mathcal F} - {\mathcal F}_0)\exp(-itM)\phi \| =0.
\end{gather}
\end{lemma}

\begin{proof} To prove (\ref{pr}), we fix $\phi(p)$ with ${\rm
supp}(\phi)\subset [r,R]$, $0<r<R$. Then we have
\begin{equation}
\| ({\mathcal F} - {\mathcal F}_0)\exp(-itM)\phi \|^2 =
\frac{1}{2\pi}\int_{-\infty}^\infty dx\left|\sum_{n=1}^NR_n(x)\int_r^Rdp\frac{e^{ixp-2it\cosh
p}}{e^p-e^{-r_n}}\phi(p)\right|^2.
\end{equation}
Now when we change variables $p\to y=\cosh p$, we see that the $p$-integrals
yield bounded functions $b_n(t,x)$
that converge to 0 as $t\to\infty$ by virtue of the Riemann--Lebesgue lemma.
By dominated convergence, this
entails that the $x$-integral over a bounded region converges to 0 for
$t\to\infty$.

To exploit this, we write
\begin{equation}\label{split}
\int_{-\infty}^\infty dx =\int_{-\infty}^{-1}dx+\int_{-1}^1 dx +\int_1^{\infty}dx.
\end{equation}
As we have just shown, the middle integral tends to 0 for $t\to\infty$. To
handle the right integral, we recall
from Part~I that $R_n(x)$ has exponential decay for $x\to\infty$,
cf.~I(2.41),~(2.42). This decay supplies the
domination we need, in combination with the pointwise convergence to 0 of
the $x$-integrand, to deduce it tends
to 0 for $t\to\infty$, too.

To handle the left integral, we use a stationary phase argument.
Specifically, we write
\begin{equation}
\exp(ixp-2it\cosh p)=(ix-2it\sinh p)^{-1}\partial_p\exp(ixp-2it\cosh p),
\end{equation}
and integrate by parts to get
\begin{equation}\label{bp}
\frac{1}{2\pi}\int_{-\infty}^{-1}dx\left|
\sum_{n=1}^NR_n(x)\int_r^Rdpe^{ixp-2it\cosh p}\partial_p\left(
\frac{1}{(x-2t\sinh p)}\frac{\phi(p)}{(e^p-e^{-r_n})}\right)\right|^2.
\end{equation}
Now we use the estimate
\begin{equation}\label{xt}
|x-2t\sinh p|^2\ge x^2+4t^2\sinh^2r,\qquad x\in(-\infty,-1],\quad p\in [r,R],\quad
 t>0,
\end{equation}
and boundedness of $R_n(x)$ on ${\mathbb R}$ to obtain an upper bound of the form
\begin{equation}\label{intl}
C\int_{-\infty}^{-1}dx\frac{1}{x^2+ct^2},\qquad C,c>0.
\end{equation}
The integrand is bounded above by the $L^1((-\infty,-1])$-function $1/x^2$
and tends to 0 as $t\to\infty$. Hence
(\ref{intl}) tends to 0 as well, so that (\ref{bp}) does, too. Therefore,
we have now proved (\ref{pr}).

In order to prove (\ref{nl}), we observe that the estimate (\ref{xt}) is
also valid for $p\in [-R,-r]$ and
$t<0$. Thus we can choose $\phi(p)$ with supp$(\phi)\subset [-R,-r]$ and
proceed in the same way as for~(\ref{pr}).

Next, we prove (\ref{nr}). To this end we recall (\ref{Wl}) and
(\ref{rhom}), which we rewrite as
\begin{equation}
{\mathcal W}(x,p)-e^{ixp}a(p)=\sum_{n=1}^N\xi_n(x)/\left(e^p-e^{-r_n}\right),
\end{equation}
with $\xi_n(x)$ admitting the two representations
\begin{gather}\label{l1}
\xi_n(x)=\left(C^{-1}\zeta\right)_n-R_n(x),\\
\label{l2}
\xi_n(x)=\left(C^{-1}D(x)R(x)\right)_n.
\end{gather}
Now for $\phi(p)$ with supp$(\phi)\subset [r,R]$ we have
\begin{gather}
\| ({\mathcal F} - {\mathcal F}_0 a(\cdot))\exp(-itM)\phi \|^2 \nonumber\\
\qquad {}=
\frac{1}{2\pi}\int_{-\infty}^\infty dx\left|\sum_{n=1}^N \xi_n(x)\int_r^Rdp\frac{e^{ixp-2it\cosh
p}}{e^p-e^{-r_n}}\phi(p)\right|^2.
\end{gather}

From (\ref{l1}) we see that $\xi_n(x)$ is bounded on ${\mathbb R}$. Using once more
the splitting (\ref{split}), it
follows that the middle integral tends to 0 as $t\to -\infty$. The right
integral can be handled by the same
stationary phase argument as before, noting that the estimate (\ref{xt}) is
also valid for $x\in [1,\infty)$ and
$t<0$.

It remains to show that the left integral tends to 0 as $t\to -\infty$. To
this end we use the second
representation (\ref{l2}) of $\xi_n(x)$. Indeed, it entails that $\xi_n(x)$
has exponential decay as
$x\to -\infty$. (Recall the definitions (\ref{defD}), (\ref{defd}).) Thus
we can once again combine the
Riemann--Lebesgue lemma and the dominated convergence theorem to deduce
convergence to 0 for $t\to -\infty$.

Finally, to prove (\ref{pl}) we can proceed in the same way as for
(\ref{nr}), noting that (\ref{xt}) also holds
for $x\in [1,\infty)$ and $p\in [-R,-r]$.
\end{proof}

We are now in the position to obtain the principal result of this section.

\setcounter{theorem}{1}

\begin{theorem}\label{theorem:3.2}
The eigenfunction transform ${\mathcal F}$ (\ref{cF}) is isometric. The strong
limits of the operator family
$\exp(it\hat{A})\exp(-it\hat{A}_0){\mathcal F}_0$ for $t\to \pm\infty$ exist and are
given by
\begin{equation}\label{Upm}
U_{\pm}={\mathcal F} {\mathcal A}_{\pm}(\cdot),
\end{equation}
where
\begin{gather}
{\mathcal A}_{+}(p)\equiv \left\{
\begin{array}{ll}
1,  &  p>0,  \\
1/a(p),  & p<0,
\end{array} \right.\\
{\mathcal A}_{-}(p)\equiv \left\{
\begin{array}{ll}
1/a(p),  &  p>0,  \\
1,  & p<0.
\end{array} \right.
\end{gather}
The $S$-operator
\begin{equation}
S_p\equiv U_{+}^{*}U_{-}
\end{equation}
equals the unitary multiplication operator
\begin{equation}
T(p)\equiv \left\{
\begin{array}{ll}
1/a(p),  &  p>0,  \\
a(p),  & p<0.
\end{array} \right.
\end{equation}
\end{theorem}

\begin{proof} We choose $\psi(p)\in C_0^{\infty}({\mathbb R}^{*})$, so that
\begin{equation}
\psi =\phi_{+} +\phi_{-},\qquad  \phi_{+}\in C_0^{\infty}((0,\infty)),\qquad
\phi_{-}\in C_0^{\infty}((-\infty,0)).
\end{equation}
By virtue of (\ref{pr}) and (\ref{pl}) we have
\begin{gather}
   \| {\mathcal F}\psi -\exp(it\hat{A})\exp(-it\hat{A}_0){\mathcal F}_0
(\phi_{+}+a(\cdot)\phi_{-})\|
   \le   \| ({\mathcal F} - \exp(it\hat{A})\exp(-it\hat{A}_0){\mathcal F}_0 )\phi_{+}\|
\nonumber\\
\qquad {}+\| ({\mathcal F}-
\exp(it\hat{A})\exp(-it\hat{A}_0){\mathcal F}_0 a(\cdot))\phi_{-}\|
      \le    \|({\mathcal F}-{\mathcal F}_0)\exp(-itM)\phi_{+}\| \nonumber\\
\qquad {}+ \|({\mathcal F}-{\mathcal F}_0
a(\cdot))\exp(-itM)\phi_{-}\|     \to 0,\qquad t\to\infty.\label{Flim}
\end{gather}

From this we deduce
\begin{gather}
\| {\mathcal F}\psi\|=\lim_{t\to\infty}\|
\exp(it\hat{A})\exp(-it\hat{A}_0){\mathcal F}_0(\phi_{+}+a(\cdot)\phi_{-})\|
\nonumber\\
\phantom{\| {\mathcal F}\psi\|}{}=\|\phi_{+}+a
(\cdot)\phi_{-}\|=\|\psi\|.
\end{gather}
Since $C_0^{\infty}({\mathbb R}^{*})$ is dense in ${\mathcal H}_p$,
it follows that ${\mathcal F}$ is
an isometry. From (\ref{Flim}) and
its analog for $t\to -\infty$ we also obtain the second assertion of the
theorem. The last assertion is then
clear from (\ref{Upm}) and isometry of ${\mathcal F}$.
\end{proof}

\section{Bound states and spectral resolution}

We begin this section by focusing on the $A$-eigenfunctions ${\mathcal W}(x,r_k+2\pi
il)$, with $x\in {\mathbb R}$, $k=1,\ldots,N$ and
$l\in{\mathbb Z}$, assuming only (\ref{r1})--(\ref{mu}). Using (\ref{W}) and
(\ref{C})--(\ref{sysN}) we see they can be
written
\begin{equation}\label{WR}
{\mathcal W}(x,r_k+2\pi il)=\exp(ir_kx-2\pi lx)d(r_k,\mu_k;x)R_k(x),\qquad l\in{\mathbb Z}.
\end{equation}
Now from (\ref{sysN}) we readily obtain
\begin{equation}
\lim_{{\rm Re}\, x\to\infty}d(r_k,\mu_k;x)R_k(x)=1,\qquad k=1,\ldots,N.
\end{equation}
For square-integrability of (\ref{WR}) near $\infty$ we should therefore
take $l\in{\mathbb N}$ when ${\rm Im}\,  r_k\in(0,\pi)$
and $l\in{\mathbb N}^{*}$ when ${\rm Im}\,  r_k\in(-\pi,0)$.

Consider now square-integrability near $-\infty$. Since $R_k(x)$ tends to
$\left(C^{-1}\zeta\right)_k\in{\mathbb C}^{*}$ for
$x\to -\infty$ (cf.~(\ref{Ras}) and I~Lemma~2.1), we see from the
definition (\ref{defd}) of $d$ that we need
$-l\in{\mathbb N}$. For ${\rm Im}\,  r_k\in(-\pi,0)$, therefore, we cannot simultaneously
have square-integrability of (\ref{WR})
near $\infty$ and near $-\infty$. In contrast, for ${\rm Im}\,  r_k\in(0,\pi)$, the
choice $l=0$ ensures
square-integrability near $\pm\infty$.

Thus far, we have not imposed restrictions on $(r,\mu)$ beyond our standing
assumptions (\ref{r1})--(\ref{mu}).
But to ensure square-integrability over ${\mathbb R}$ of ${\mathcal W}(x,r_k)$ for ${\rm Im}\,
r_k\in(0,\pi)$, we should obviously
require absence of poles for real $x$ (the assumption made in Lemma~\ref{lemma:2.1}).
Doing so, we obtain $A$-eigenfunctions
whose restrictions to the real axis are in ${\mathcal H}_x$. We can only prove
pairwise orthogonality of these functions,
however, when we make the same assumptions as in Lemma~\ref{lemma:2.2}.

\begin{lemma}\label{lemma:4.1}
With the assumption of Lemma~\ref{lemma:2.1} in force,  the functions
\begin{equation}\label{psin}
\psi_n(x)\equiv {\mathcal W}(x,r_n)=\mu_n(x)\exp(-ir_nx)R_n(x),\qquad n=1,\ldots,N_{+},
\end{equation}
satisfy
\begin{equation}\label{as+}
\psi_n(x)\sim \exp(ir_nx),\qquad {\rm Re}\, x\to\infty,
\end{equation}
\begin{equation}\label{as-}
\psi_n(x)\sim c_n(C^{-1}\zeta)_n\exp(-ir_nx),\qquad {\rm Re}\, x\to -\infty,
\end{equation}
uniformly for ${\rm Im}\,  x$ in compacts, and their restrictions to ${\mathbb R}$ belong to
${\mathcal H}_x$. Now suppose that the
requirements of Lemma~\ref{lemma:2.2} are met. Then the functions
$\psi_1(x),\ldots,\psi_{N_{+}}(x)$ are pairwise orthogonal in ${\mathcal H}_x$.
\end{lemma}

\begin{proof} We have already shown square-integrability and the asymptotics
(\ref{as+})--(\ref{as-}). To prove
pairwise orthogonality, we invoke the eigenvalue equations
\begin{equation}\label{AWn}
(A{\mathcal W})(x,r_n)=2\cosh(r_n){\mathcal W}(x,r_n),\qquad n=1,\ldots,N_{+}.
\end{equation}

From (\ref{r1}), (\ref{r2}) and (\ref{lan}), we see that the eigenvalues
are real and distinct. Letting $n\ne
m$, we now use (\ref{AWn}) to write
\begin{gather}
2[\cosh r_m-\cosh r_n ](\psi_n,\psi_m)    \nonumber\\
\qquad {}=   \int_{-\infty}^\infty
dx(\psi_n^{*}(x)[\psi_m(x-i)+V_a(x)\psi_m(x+i)
+V_b(x)\psi_m(x)]
\nonumber\\
   \label{evdiff}
\qquad {}-[\psi_n^{*}(x+i)+V_a^{*}(x)\psi_n^{*}(x-i)+V_b^{*}(x)\psi_n^{*}(x)]\psi_m(x)).
\end{gather}
Using (\ref{fsa}), we can write the rhs as
\begin{equation}\label{Jint}
\int_{-\infty}^\infty dx (J_{nm}(x)-J_{nm}(x+i)),
\end{equation}
with
\begin{equation}
J_{nm}(x)\equiv \psi_n^{*}(x)\psi_m(x-i)-V_a(x-i)\psi_n^{*}(x-i)\psi_m(x).
\end{equation}
We now recall (\ref{RE}). It entails we may write
\begin{equation}
\psi_n(x)=\mu_n\exp(-ir_nx)E_n(x)/\tau(x),\qquad n=1,\ldots,N_{+},
\end{equation}
with $E_n(x)$ entire. Using also (\ref{Vn}) and (\ref{tst}), we deduce that
$J_{nm}(x)$ can be rewritten as
\begin{gather}
J_{nm}(x)    =    \overline{\mu_n}\mu_m \left( \exp(-ir_nx-ir_m(x-i))
\frac{E_n^{*}(x)E_m(x-i)}{\tau(x-i)^2}\right.
\nonumber \\
\phantom{J_{nm}(x)    =}{}  - \left. \exp(-ir_n(x-i)-ir_mx)
\frac{E_n^{*}(x-i)E_m(x)}{\tau(x-i)^2}\right).
\end{gather}
This representation shows that $J_{nm}(x)$ has no poles on and inside the
contour $\Gamma$ defined below
(\ref{Jxpq}) in the proof of Lemma~\ref{lemma:2.2}. Thus we can use the same reasoning
as in that proof to infer that~(\ref{Jint}) vanishes.
 (The vanishing of the boundary terms is here a
simple consequence of the asymptotics
(\ref{as+}), (\ref{as-}) and (\ref{Vas}).) Hence pairwise orthogonality
follows from vanishing of the lhs of~(\ref{evdiff}).
\end{proof}

The lemma just proved together with our next lemma are the key to
clarifying the character of the subspace ${\rm
Ran}({\mathcal F})^{\perp}$, when the assumptions of Lemma~\ref{lemma:2.2} are satisfied. From
Theorem~\ref{theorem:3.2} we already know that in
that case ${\mathcal F}$ is an isometry, so that we have
\begin{equation}
{\mathcal F}^{*}{\mathcal F} ={\bf 1},\qquad {\mathcal F}{\mathcal F}^{*} ={\bf 1}-P,
\end{equation}
with $P$ the projection on ${\rm
Ran}({\mathcal F})^{\perp}$. In the next lemma we obtain a formula for $({\mathcal F}^{*}
f_1,{\mathcal F}^{*} f_2)\!$ with $f_1,f_2\in C_0^\infty({\mathbb R})$,
from which this projection can be explicitly obtained. To prove the
pertinent formula, however, we need only
impose our second requirement (\ref{lan}), together with absence of poles
for real~$x$, cf.~Lemma~\ref{lemma:2.1}.

\begin{lemma}\label{lemma:4.2}
Assume that the spectral data satisfy (\ref{lan}), and assume $R(x)$ has no
real poles. Then we have for all
$f_1,f_2\in C_0^\infty({\mathbb R})$
\begin{equation}\label{key}
({\mathcal F}^{*} f_1,{\mathcal F}^{*} f_2)=(f_1,f_2)-\sum_{n=1}^{N_{+}}
\nu_n(f_1,\psi_n)(\psi_n,f_2),
\end{equation}
where $\psi_n$ is defined by (\ref{psin}).
\end{lemma}

\begin{proof} Our starting point is the formula
\begin{equation}\label{stf}
({\mathcal F}^{*} f_1,{\mathcal F}^{*} f_2)=\frac{1}{2\pi}\lim_{R\to\infty}\int_{-\infty}^\infty
dy\overline{f_1(y)}\int_{-\infty}^\infty dxf_2(x){\mathcal I}_R(x,y),
\end{equation}
where
\begin{equation}
{\mathcal I}_R(x,y)\equiv
\int_{-R}^Rdp{\mathcal W}^{*}(x,p){\mathcal W}(y,p),\qquad
x,y\in{\mathbb R}.
\end{equation}
We are going to exploit that ${\mathcal W}(x,p)$ is the product of the plane wave
$\exp(ixp)$ and a~function of $p$ that
is meromorphic and $2\pi i$-periodic, cf.~(\ref{W}).

For this purpose we define the rectangular contour $C$ connecting
$-R$, $R$, $R+2\pi i$ and $-R+2\pi i$ in the
$p$-plane, and put
\begin{equation}
{\mathcal I}_C(x,y)\equiv \oint_C  dp{\mathcal W}^{*}(x,p){\mathcal W}(y,p),
\qquad x,y\in{\mathbb R}.
\end{equation}
(Of course, the $*$ refers here to the $p$-dependence, cf.~(\ref{fconj}).)
Due to our requirements on
$r_1,\ldots,r_N$, the only singularities of the integrand  inside $C$
consist of simple poles at the $2N$
distinct points
\begin{gather}\label{p+}
p=r_n,2\pi i-r_n,\qquad n=1,\ldots,N_{+},
\\
\label{p-}
p=-r_n,2\pi i+r_n,\qquad n=N_{+}+1,\ldots,N,
\end{gather}
on the imaginary axis. Hence Cauchy's theorem yields
\begin{equation}
{\mathcal I}_C(x,y)=2\pi i{\mathcal R}(x,y),
\end{equation}
where ${\mathcal R}(x,y)$ denotes the sum of the residues.

On the other hand, we can also write
\begin{equation}\label{IB}
{\mathcal I}_C(x,y)={\mathcal I}_R(x,y)-e^{2\pi(x-y)}{\mathcal I}_R(x,y)+{\mathcal B}_R(x,y),
\end{equation}
with
\begin{equation}\label{BR}
{\mathcal B}_R(x,y)\equiv \left( \int_R^{R+2\pi i}+\int_{-R+2\pi i}^{-R}\right)
dp{\mathcal W}^{*}(x,p){\mathcal W}(y,p).
\end{equation}
Hence we have
\begin{equation}\label{IRB}
{\mathcal I}_R(x,y)=\left(1-e^{2\pi(x-y)}\right)^{-1}
[2\pi i{\mathcal R}(x,y)-{\mathcal B}_R(x,y)].
\end{equation}

We proceed to calculate the contribution of the residue sum to the inner
product
\begin{gather}
\left({\mathcal F}^{*} f_1,{\mathcal F}^{*} f_2\right)
 =    \frac{1}{2\pi}\lim_{R\to\infty}\int_{-\infty}^\infty
dy\overline{f_1(y)}\int_{-\infty}^\infty dxf_2(x)
\nonumber\\
\phantom{\left({\mathcal F}^{*} f_1,{\mathcal F}^{*} f_2\right)  =}{}
\times \left(1-e^{2\pi(x-y)}\right)^{-1}[2\pi i{\mathcal R}(x,y)-{\mathcal B}_R(x,y)],\label{IP1}
\end{gather}
where we combined (\ref{stf}) and (\ref{IRB}). First, we show that the
residues at the poles (\ref{p-}) cancel
pairwise. Indeed, fixing $l\in\{ N_{+}+1,\ldots,N\}$, the residue sum at
$p=-r_l,2\pi i+r_l$ of the pertinent
function
\begin{equation}\label{pert}
\exp[ip(y-x)]\left( 1-\sum_{k=1}^N\frac{R_k^{*}(x)}{e^p-e^{r_k}}\right)
\left( 1-\sum_{m=1}^N\frac{R_m(y)}{e^p-e^{-r_m}}\right)
\end{equation}
equals
\begin{gather}
\exp[-ir_l(y-x)]\left( 1-\sum_k\frac{R_k^{*}(x)}{e^{-r_l}-e^{r_k}}\right)
\left( -\frac{R_l(y)}{e^{-r_l}}\right)
\nonumber\\
\qquad{} +\exp[(ir_l-2\pi)(y-x)]\left( -\frac{R_l^{*}(x)}{e^{r_l}}\right) \left(
1-\sum_m\frac{R_m(y)}{e^{r_l}-e^{-r_m}}\right).
\end{gather}
Recalling the system (\ref{sysN}), we see that this equals
\begin{gather}
 -\exp[-ir_l(y-x)]d^{*}(r_l,\mu_l;x)R_l^{*}(x)e^{r_l}R_l(y)
\nonumber\\
  \qquad  -\exp[(ir_l-2\pi)(y-x)]e^{-r_l}R_l^{*}(x)d(r_l,\mu_l;y)R_l(y).
\end{gather}

From the definition (\ref{defd}) of $d$ we see that this is proportional to
\begin{gather}
 \exp[-ir_l(y-x)]\overline{\mu_l}\exp[-2i(r_l+i\pi)x]e^{r_l}
\nonumber\\
\qquad{}+\exp[(ir_l-2\pi)(y-x)]e^{-r_l}\mu_l\exp[-2i(r_l+i\pi)y]
\nonumber\\
\qquad{}  =\exp[-ir_l(y+x)]e^{2\pi x}\left(\overline{\mu_l}e^{r_l}+\mu_le^{-r_l}\right),\label{resvan}
\end{gather}
which vanishes due to (\ref{lan}).

We are therefore left with the residues of (\ref{pert}) at the points
(\ref{p+}). The residue sum at $p=r_n,2\pi
i-r_n$ for
$n\in\{ 1,\ldots,N_{+}\}$ equals (using (\ref{sysN}), (\ref{lan}) and
(\ref{psin}))
\begin{gather}
 \exp[ir_n(y-x)]\left(-\frac{R_n^{*}(x)}{e^{r_n}}\right)
\mu_n e^{-2ir_ny}R_n(y)
\nonumber\\
\qquad{}+\exp[(-ir_n-2\pi)(y-x)]\overline{\mu_n}e^{-2ir_nx}R_n^{*}(x)\left(-\frac{R_n(y)
}{e^{-r_n}}\right)
\nonumber\\
\qquad{} =-\exp[-ir_n(y+x)]R_n^{*}(x)R_n(y)\left(\mu_ne^{-r_n}+e^{2\pi
(x-y)}\overline{\mu_n}e^{r_n}\right)
\nonumber\\
\qquad{} =i\nu_n^{-1} \left(1-e^{2\pi(x-y)}\right)e^{-ir_nx}R_n^{*}(x)e^{-ir_ny}R_n(y)
\nonumber\\
\qquad =i\nu_n \left(1-e^{2\pi(x-y)}\right)\overline{\psi_n(x)}\psi_n(y).
\end{gather}
Substituting this in (\ref{IP1}) and comparing to (\ref{key}), we deduce
that it remains to prove
\begin{equation}\label{IP2}
\frac{1}{2\pi}\lim_{R\to\infty}\int_{-\infty}^\infty dy\overline{f_1(y)}\int_{-\infty}^\infty
dxf_2(x)\frac{{\mathcal B}_R(x,y)}{e^{2\pi(x-y)}-1}=(f_1,f_2).
\end{equation}

To this end we first rewrite ${\mathcal B}_R(x,y)$ (\ref{BR}) as
\begin{gather}
{\mathcal B}_R(x,y)   =  \int_{R-i\pi }^{R+i\pi}
dp[{\mathcal W}^{*}(x,p+i\pi)
{\mathcal W}(y,p+i\pi)-{\mathcal W}^{*}(x,-p+i\pi){\mathcal W}(y,-p+i\pi)]
\nonumber\\
\phantom{{\mathcal B}_R(x,y) }{}  =    e^{\pi(x-y)}\int_{R-i\pi }^{R+i\pi}
dp\left[e^{ip(y-x)}A(p,x,y)-e^{ip(x-y)}A(-p,x,y)\right],\label{BA}
\end{gather}
where the auxiliary function $A$ is given by
\begin{equation}
A(p,x,y)\equiv \left( 1+\sum_n\frac{R_n^{*}(x)}{e^p+e^{r_n}}\right)
\left( 1+\sum_m\frac{R_m(y)}{e^p+e^{-r_m}}\right).
\end{equation}
We now claim that the identity
\begin{equation}\label{Ap}
A(p,x,x)=A(-p,x,x)
\end{equation}
holds true.

To prove (\ref{Ap}), we observe that the functions $A(\pm p,x,x)$ are $2\pi
i$-periodic in $p$ and bounded for
$|{\rm Re}\, p|\to\infty$, and that they have simple poles in the period strip
${\rm Im}\,  p\in[0,2\pi]$ at $p=i\pi \pm
r_n$, $n=1,\ldots,N$. Using (\ref{sysN}) and (\ref{defd}) in the same way as above
(cf.~(\ref{pert})--(\ref{resvan})), we readily verify that the functions
have equal residues. By Liouville's
theorem, it now follows that $A(p,x,x)-A(-p,x,x)$ does not depend on $p$.

To show that this difference vanishes, we need only compare the ${\rm Re}\,
p\to\infty$ limits of $A(\pm p,x,x)$.
Obviously, we have
\begin{equation}
\lim_{{\rm Re}\, p\to\infty}A(p,x,x)=1,\qquad \lim_{{\rm Re}\,
p\to\infty}A(-p,x,x)=\lambda^{*}(x)\lambda(x),
\end{equation}
where
\begin{equation}\label{deflam}
\lambda(x)\equiv 1+\sum_{m=1}^Ne^{r_m}R_m(x).
\end{equation}
Now from I(D.17) we see that (\ref{lan}) entails
\begin{equation}\label{laid}
\lambda^{*}(x)\lambda(x)=1.
\end{equation}
Hence our claim (\ref{Ap}) follows.

Next, we introduce functions $B$ and $C$ by setting
\begin{gather}\label{Bd}
A(p,x,y)=1+e^{-p}B(p,x,y),
\\
\label{Cd}
A(-p,x,y)=\lambda^{*}(x)\lambda(y)+e^{-p}C(p,x,y).
\end{gather}
Due to (\ref{Ap}), these functions are related by
\begin{equation}\label{BC}
B(p,x,x)=C(p,x,x).
\end{equation}
We now rewrite (\ref{BA}) as
\begin{equation}
{\mathcal B}_R(x,y)=
e^{\pi(x-y)}\left( {\mathcal B}_R^{(d)}(x,y)+{\mathcal B}_R^{(r)}(x,y)\right),
\end{equation}
with
\begin{gather}\label{BRd}
{\mathcal B}_R^{(d)}(x,y)\equiv \int_{R-i\pi }^{R+i\pi}
dp\left[e^{ip(y-x)}-e^{ip(x-y)}\lambda^{*}(x)\lambda(y)\right],
\\
\label{BRr}
{\mathcal B}_R^{(r)}(x,y)\equiv \int_{R-i\pi }^{R+i\pi}
dpe^{-p}\left[e^{ip(y-x)}B(p,x,y)-e^{ip(x-y)}C(p,x,y)\right].
\end{gather}
The point is that we can get rid of the remainder term ${\mathcal B}_R^{(r)}(x,y)$
by using (\ref{BC}).

Specifically, (\ref{BC}) entails we may write
\begin{equation}
{\mathcal B}_R^{(r)}(x,y)=\int_{R-i\pi }^{R+i\pi} dpe^{-p}\int_x^yds\partial_s\left(
e^{ip(s-x)}B(p,x,s)-e^{ip(x-s)}C(p,x,s)\right).
\end{equation}
Now from the definitions (\ref{Bd}), (\ref{Cd}) of $B$ and $C$ we deduce
that for all $s$-values between~$x$ and
$y$ we have
\begin{gather}
\left| \partial_s\left(
e^{ip(s-x)}B(p,x,s)-e^{ip(x-s)}C(p,x,s)\right)\right| \le |p|D(x,y),\nonumber\\
\qquad {\rm Re}\, p\ge R,\qquad  |{\rm Im}\,  p|\le \pi,
\end{gather}
where $D(x,y)$ is a positive function that is bounded for $x,y$ varying
over ${\mathbb R}$-compacts. Thus we obtain
\begin{equation}
\left|{\mathcal B}_R^{(r)}(x,y)\right|\le 2\pi e^{-R}\left(R^2+\pi^2\right)^{1/2}|y-x|D(x,y).
\end{equation}
Hence the contribution of ${\mathcal B}_R^{(r)}(x,y)$ to the lhs of (\ref{IP2})
vanishes.

We are now reduced to showing
\begin{equation}\label{IP3}
\lim_{R\to\infty}\int_{-\infty}^\infty dy\overline{f_1(y)}\int_{-\infty}^\infty
dxf_2(x)\frac{{\mathcal B}_R^{(d)}(x,y)}{\sinh (\pi(x-y))}=4\pi (f_1,f_2).
\end{equation}
Calculating the integral (\ref{BRd}), we can write the result as
\begin{equation}
{\mathcal B}_R^{(d)}(x,y)=2\sinh(\pi(x-y))[C_c(R;x,y)+C_s(R;x,y)],
\end{equation}
where
\begin{gather}
C_c(R;x,y)\equiv \left(  \frac{1-\lambda^{*}(x)\lambda(y)}{i(y-x)}\right)
\cos (y-x)R,
\\
C_s(R;x,y)\equiv (1+\lambda^{*}(x)\lambda(y))\frac{\sin (y-x)R}{y-x}.
\end{gather}
By virtue of (\ref{laid}) and the Riemann--Lebesgue lemma, the contribution
of $C_c(R;x,y)$ to~(\ref{IP3})
vanishes. Using the tempered distribution limit
\begin{equation}
\lim_{c\to\infty}\frac{\sin cx}{x}=\pi \delta(x)
\end{equation}
and (\ref{laid}), the remaining term $C_s(R;x,y)$ yields (\ref{IP3}).
\end{proof}

Clearly, we can rewrite (\ref{key}) as the operator identity
\begin{equation}\label{Pfor}
{\mathcal F}{\mathcal F}^{*}={\bf 1}-\sum_{n=1}^{N_{+}}\nu_n\psi_n\otimes \overline{\psi_n},
\end{equation}
where the rank-one operator $\psi\otimes \chi$, with $\psi,\chi\in{\mathcal H}_x$,
is defined by
\begin{equation}
(\psi\otimes\chi)f=(\overline{\chi},f)\psi,\qquad f\in{\mathcal H}_x.
\end{equation}

For the remainder of this section, we assume that the requirements of
Lemma~\ref{lemma:2.2} are met. Then ${\mathcal F}$ is an
isometry (as proved in Theorem~\ref{theorem:3.2}), and the functions
$\psi_1,\ldots\!,\psi_{N_{+}}\!$~(\ref{psin}) are pairwise
orthogonal in ${\mathcal H}_x$ (as proved in Lemma~\ref{lemma:4.1}). In view of (\ref{Pfor}),
the projection~$P$ on
Ran$({\mathcal F})^\perp$ can be written
\begin{equation}\label{Pid}
P=\sum_{n=1}^{N_{+}}\nu_n\psi_n\otimes \overline{\psi_n}.
\end{equation}
In particular, this entails the norm formula
\begin{equation}
(\psi_n,\psi_n)=\nu_n^{-1},\qquad n=1,\ldots,N_{+}.
\end{equation}

We are now in the position to turn the provisional definition of the
self-adjoint Hilbert space operator
$\hat{A}$ (see the end of Section~2) into a final one: We define $\hat{A}$
on Ran$({\mathcal F})^\perp$ by (linear
extension of)
\begin{equation}\label{Afin}
\hat{A}\psi_n\equiv 2\cosh(r_n)\psi_n,\qquad n=1,\ldots,N_{+}.
\end{equation}

\setcounter{theorem}{2}
\begin{theorem}\label{theorem:4.3}
The operator $\hat{A}$ is essentially self-adjoint on the dense subspace
\begin{equation}\label{core}
{\mathcal C} \equiv {\mathcal P} \oplus {\rm Span}(\psi_1,\ldots,\psi_{N_{+}}),
\end{equation}
and its action on ${\mathcal C}$ coincides with that of the A$\Delta$O $A$. The
operator $\hat{A}$ has absolutely continuous
spectrum $[2,\infty)$ with multiplicity two, and point spectrum $\{
2\cosh(r_1),\ldots,$ $2\cosh(r_{N_{+}})\}$ with multiplicity one.
\end{theorem}

\begin{proof} The first assertion follows by combining Lemmas~\ref{lemma:2.1}
and~\ref{lemma:2.2}
with (\ref{psin}), (\ref{AWn}) and~(\ref{Afin}).
 Denoting the domain of $\hat{A}$ by ${\mathcal D}$, the restriction of
$\hat{A}$ to ${\mathcal D}\cap{\rm Ran}({\mathcal F})$
is unitarily equivalent to multiplication by $2\cosh(p)$ on ${\mathcal H}_p$,
cf.~(\ref{hA}). Together with the
distinctness of the numbers $r_1,\ldots,r_{N_{+}}\in i(0,\pi)$, this
entails the second assertion.
\end{proof}

We conclude this section with some further observations concerning three
special cases. Taking first $N_{-}=0$,
we recall that the assumptions
\begin{equation}
0<-ir_1<\cdots <-ir_N<\pi,\qquad \nu_n\in(0,\infty),\qquad n=1,\ldots,N,
\end{equation}
are sufficient for all of our results to hold true, cf.~Lemma~\ref{lemma:A.1}. Now in
this special case the A$\Delta$O~$A$
satisfies
\begin{equation}
A=S_{+}^2-2,
\end{equation}
where $S_{+}$ is the A$\Delta$O
\begin{gather}
S_{+}\equiv \exp(-i\partial_x/2)+V(x)\exp(i\partial_x/2),
\\
\label{defV}
V(x)\equiv \lambda(x)/\lambda(x+i/2),
\end{gather}
with $\lambda(x)$ given by (\ref{deflam}); moreover,
\begin{equation}
(S_{+}{\mathcal W})(x,p)=\left(e^{p/2}+e^{-p/2}\right){\mathcal W}(x,p).
\end{equation}
(These assertions follow from I~Theorem~3.3.) Using the above results, we
can associate a self-adjoint operator
$\hat{S}_{+}$ to $S_{+}$ by setting
\begin{gather}
\hat{S}_{+}{\mathcal F} f\equiv {\mathcal F} M_{+}f,
\qquad (M_{+}f)(p)\equiv 2\cosh(p/2)f(p),\qquad
f\in{\mathcal D}(M_{+}),
\\
\hat{S}_{+}\psi_n\equiv 2\cosh(r_n/2)\psi_n,\qquad n=1,\ldots,N_{+}.
\end{gather}
It follows just as for $A$ that the dense subspace  ${\mathcal C}$ (\ref{core}) is a
core for $\hat{S}_{+}$ on which the
$\hat{S}_{+}$-action coincides with that of $S_{+}$.

Secondly, we consider the special case $N=2M$, $N_{+}=N_{-}=M$, together with
spectral data
\begin{gather}
0<-ir_1<\cdots<-ir_M<\pi/2,\qquad
 \nu_j\in(0,\infty),\nonumber\\
r_{N-j+1}=r_j-i\pi,\qquad
\nu_{N-j+1}=\nu_j,
\end{gather}
where
$j=1,\ldots,M$. Again, this suffices for all of the above Hilbert space
results to be valid, cf.~Lemma~\ref{lemma:A.2}. In
terms of the particle variables defined in Appendix~A, this choice of
spectral data amounts to
\begin{equation}
0<q_M^{+}<\cdots<q_1^{+},\!\qquad q_j^{-}=-q_{M-j+1}^{+},\!\qquad
\theta_j^{-}=\theta_{M-j+1}^{+},\!\qquad j=1,\ldots,M.\!
\end{equation}
Its distinguishing feature consists in the potential $V_b(x)$ being
identically zero. Moreover, after taking
$M\to N$ and performing a scaling $x,p\to 2x,p/2$, the class of A$\Delta$Os~$A$,
together with their reflectionless
eigenfunctions and associated self-adjoint operators $\hat{A}$, amounts to
the class of A$\Delta$Os $S_{+}$,
together with their reflectionless eigenfunctions and associated
self-adjoint operators $\hat{S}_{+}$.
Once more, this follows from I~Theorem~3.3. (The ordering we used there is
different, but this is
inconsequential. Indeed, all of the pertinent quantities are permutation
invariant.)

Thirdly, we consider the special case $N_{+}=0$. From
Lemma~\ref{lemma:A.1} we then infer that the assumptions
\begin{equation}
-\pi<-ir_N<\cdots<-ir_1<0,\qquad \nu_n\in (0,\infty),\qquad n=1,\ldots,N,
\end{equation}
suffice for the validity of our Hilbert space results. In this case we have
the A$\Delta$O identity
\begin{equation}
A=S_{-}^2+2,
\end{equation}
where
\begin{equation}
S_{-}\equiv \exp(-i\partial_x/2)-V(x)\exp(i\partial_x/2),
\end{equation}
and $V(x)$ is again given by (\ref{defV}); moreover,
\begin{equation}
(S_{-}{\mathcal W})(x,p)=\left(e^{p/2}-e^{-p/2}\right){\mathcal W}(x,p).
\end{equation}
(These assertions are also a consequence of I~Theorem~3.3.) Since ${\mathcal F}$ is
unitary when $N_{+}$ vanishes, we can
define a self-adjoint operator $\hat{S}_{-}$ by
\begin{equation}
\hat{S}_{-}\equiv {\mathcal F} M_{-}{\mathcal F}^{*},\qquad
 (M_{-}f)(p)\equiv 2\sinh(p/2)f(p),\qquad
 f\in{\mathcal D}(M_{-}).
\end{equation}
As before, the subspace ${\mathcal C}={\mathcal P}$ is a core for $\hat{S}_{-}$, on which the
$\hat{S}_{-}$-action coincides with
that of $S_{-}$.

We would like to point out that the absence of bound states for
the case $N_{+}=0$ is a~quite remarkable feature. Indeed, for
reflectionless self-adjoint Schr\"odinger and Jaco\-bi operators,
absence of bound states implies that the potentials are trivial
(constant), whereas here one obtains an infinite-dimensional
family of nontrivial potential pairs~$V_a$,~$V_b$. When one takes
the time dependence introduced in Part~II into account, this
family of reflectionless self-adjoint operators without bound
states yields the left-moving soliton solutions to the analytic
version of the Toda lattice studied in Part~II.

We can also use the $N_{+}=0$ special case to illustrate the ambiguity
issue discussed in the introduction,
cf.~in particular the paragraph below (\ref{cF}). Let us begin by noting
that all of the wave functions
${\mathcal W}(x,p)$ studied in this series of papers satisfy
\begin{equation}
(D{\mathcal W})(x,p)=2\cosh(2\pi x){\mathcal W}(x,p),
\end{equation}
where the dual A$\Delta$O $D$ is given by
\begin{equation}
D\equiv \exp(2\pi i\partial_p)+\exp(-2\pi i\partial_p).
\end{equation}
Indeed, this is plain from ${\mathcal W}(x,p)$ being the product of the factor
$\exp(ixp)$ and a factor that is a rational
function of $\exp(p)$, cf.~(\ref{W}). For $N_{+}=0$ the operator ${\mathcal F}$ is
unitary, so we can define a
self-adjoint operator $\hat{D}$ on ${\mathcal H}_p$ by setting
\begin{equation}
\hat{D}\equiv {\mathcal F}^{*}M_x{\mathcal F},\qquad (M_xf)(x)\equiv 2\cosh(2\pi x)f(x),\qquad
f\in{\mathcal D}(M_x).
\end{equation}
The action of $\hat{D}$ on the
core ${\mathcal F}^{*}\left(C_0^{\infty}({\mathbb R})\right)$ now coincides with the action of
the A$\Delta$O $D$. (This follows in the
same way as the analogous assertion in Lemma~\ref{lemma:2.2}.)

The upshot is that we have associated to the free A$\Delta$O $D$ an
infinite-dimensional family of distinct
self-adjoint reflectionless operators $\hat{D}$ without bound states.
(Indeed, the function $\lambda(x)$
(\ref{deflam}) plays the same role for $\hat{D}$ as the function $a(p)$
(\ref{af}) plays for $\hat{A}$.)
Interchanging $x$ and $p$ and performing a scaling by $2\pi$, we see that
we obtain a similar family associated
with the free A$\Delta$O $A_0$ (\ref{A0}), as announced.

Finally, we would like to use $D$ with $N_{+}>0$ to illustrate that Hilbert
space operators
associated to A$\Delta$Os may look symmetric at first sight, even when they
are not symmetric. Indeed, the symmetry
property is far more elusive than may be apparent from the above results.
(For instance, symmetry is
probably generically violated when the requirements of Lemma~\ref{lemma:2.2} are not
met, cf.~also our results in
Ref.~\cite{hilb}.)

For this purpose we observe that whenever the assumption of Lemma~\ref{lemma:2.1} is
satisfied, we may define an operator
$\hat{D}$ on the subspace ${\mathcal F}^{*}\left(C_0^{\infty}({\mathbb R})\right)$ via
\begin{equation}
\hat{D}{\mathcal F}^{*}f\equiv {\mathcal F}^{*}M_xf,\qquad
 f\in C_0^{\infty}({\mathbb R}).
\end{equation}
(The point is that we have
\begin{equation}
{\rm Ker}({\mathcal F}^{*})\cap C_0^{\infty}({\mathbb R})=\{ 0\},
\end{equation}
by the argument proving (\ref{KerF});
 hence $\hat{D}$ is well defined.) Now with the stronger requirements of
Lemma~\ref{lemma:2.2} in effect, ${\mathcal F}$ is isometric, so that
${\mathcal F}^{*}\left(C_0^{\infty}({\mathbb R})\right)$
is dense in ${\mathcal H}_p$. For $N_{+}>0$, however, the densely
defined operator $\hat{D}$ is {\em not} symmetric.

This assertion can be verified in two ways, both of which are illuminating.
First, we can proceed as in the
proof of Lemma~\ref{lemma:2.2} to try and show symmetry. Doing so, we are led to
investigate the residue sum of the function
(\ref{pert}) in the strip ${\rm Im}\,  p\in[0,2\pi]$. As we have seen below
(\ref{pert}), this residue sum vanishes only
when $N_{+}=0$, so that $\hat{D}$ is not symmetric for $N_{+}>0$.

The second way in which symmetry violation for $N_{+}>0$ can be established
hinges on the finite-dimensionality
of Ran$({\mathcal F})^{\perp}$ already detailed above. Specifically,
Ran$({\mathcal F})^\perp$ is $N_{+}$-dimensional, and this
suffices to rule out symmetry for $N_{+}>0$.

To see this, assume $\hat{D}$ is
symmetric on ${\mathcal F}^{*}\left(C_0^{\infty}({\mathbb R})\right)$. Then it follows as
before that $\hat{D}$ is essentially
self-adjoint on ${\mathcal F}^{*}\left(C_0^{\infty}({\mathbb R})\right)$, and also that we have
\begin{equation}
\exp(it\hat{D}){\mathcal F}^{*}f={\mathcal F}^{*}\exp(itM_x)f,\qquad
 f\in C_0^{\infty}({\mathbb R}),\qquad t\in{\mathbb C}.
\end{equation}
Choosing $t$ real and taking closures, this entails
\begin{equation}
\exp(it\overline{\hat{D}}){\mathcal F}^{*}={\mathcal F}^{*}\exp(itM_x),
\end{equation}
and so
\begin{equation}
\exp(it\overline{\hat{D}})={\mathcal F}^{*}\exp(itM_x){\mathcal F},\qquad t\in{\mathbb R},
\end{equation}
by isometry of ${\mathcal F}$. From this we deduce that the unitary one-parameter
group $\exp(itM_x)$ leaves Ran$({\mathcal F})$
invariant. Hence it leaves Ran$({\mathcal F})^\perp$ invariant, too. Since the
generator $M_x$ has solely continuous
spectrum and Ran$({\mathcal F})^\perp$ is $N_{+}$-dimensional, we must have
$N_{+}=0$, as advertized.


\renewcommand{\thesection}{A}
\setcounter{equation}{0}
\setcounter{lemma}{0}

\section*{Appendix A. Absence
of $\boldsymbol{\tau(x)}$-zeros for $\boldsymbol{{\rm Im}\,  x\in [-1,0]}$}

In this appendix we state and prove two lemmas that have a bearing on the
most restrictive (fourth) requirement
made in Section~2. We recall that this requirement consists in the
restrictions (\ref{lan})--(\ref{lapo}) on
the spectral data $(r,\mu(x))$ given by the first paragraph of Section~2,
and in the additional restriction that
$\tau(x)$ (\ref{tau}) have no zeros for ${\rm Im}\,  x\in[-1,0]$. Our first lemma
shows that when $N_{+}$ or $N_{-}$
vanishes, there is no need for the latter restriction.

\begin{lemma}\label{lemma:A.1}
Assume that the spectral data satisfy
\begin{equation}
0<-ir_1<\cdots <-ir_N<\pi,\qquad
 \nu_n\in(0,\infty),\qquad n=1,\ldots,N,
\end{equation}
or
\begin{equation}
-\pi <-ir_N<\cdots <-ir_1<0,\qquad \nu_n\in(0,\infty),\qquad n=1,\ldots,N,
\end{equation}
where $\nu_n$ is given by (\ref{lan}). Then $\tau(x)$ does not vanish in
the strip ${\rm Im}\,  x\in[-1,0]$.
\end{lemma}

For $N_{+}N_{-}>0$, however, it seems likely that zeros of $\tau(x)$ in the
critical strip are not excluded by
(\ref{lan})--(\ref{lapo}). (In particular, for $N_{+}=N_{-}=1$, we can show
that $\tau(x)$ may have zeros at
the strip boundaries.) But an extra restriction on $r$ ensures absence of
critical zeros.

\begin{lemma}\label{lemma:A.2}
Assume that the spectral data satisfy $N_{+}N_{-}>0$ and that in addition
to (\ref{lan})--(\ref{lapo}) we have
\begin{equation}\label{ass2}
0<-ir_1<\cdots <-ir_{N_{+}}\le \frac{\pi}{2},\qquad -\pi
<-ir_N<\cdots<-ir_{N_{+}+1}\le -\frac{\pi}{2}.
\end{equation}
Then $\tau(x)$ has no zeros for ${\rm Im}\,  x\in[-1,0]$.
\end{lemma}

To prove these lemmas we invoke Section~5 in Part~II. To this end, we
require from now on
(\ref{lan})--(\ref{lapo}). Then we put
\begin{equation}
\alpha_j^{+}\equiv -ir_j,\qquad j=1,\ldots,N_{+},\qquad
\alpha_l^{-}\equiv -ir_{N_{+}+l}+\pi,\qquad l=1,\ldots,N_{-},
\end{equation}
so that
\begin{equation}\label{albig}
\alpha_n^{\delta}\in(0,\pi),\qquad n=1,\ldots,N_{\delta},\qquad
\delta =+,-.
\end{equation}
Next, we define real numbers
\begin{gather}
q^{+}_j\equiv \ln (\cot (\alpha^{+}_j/2)),\qquad j=1,\ldots,N_{+},\nonumber\\
q_l^{-}\equiv -\ln (\cot(\alpha_l^{-}/2)),\qquad l=1,\ldots,N_{-},
\end{gather}
and positive `potentials'
\begin{gather}
V_j^{+}(q)\equiv \prod_{1\le k\le N_{+},k\ne j}|\coth
[(q_j^{+}-q_k^{+})/2]|\nonumber\\
\phantom{V_j^{+}(q)\equiv}{}\times \prod_{1\le l\le N_{-}}|\tanh [(q_j^{+}-q_l^{-})/2]|,
\qquad j=1,\ldots,N_{+},\label{V+}
\\
\label{V-}
V_l^{-}(q)\equiv \prod_{1\le m\le N_{-},m\ne l}|\coth
[(q_l^{-}-q_m^{-})/2]|\nonumber\\
\phantom{V_j^{-}(q)\equiv}{}\times
\prod_{1\le j\le N_{+}}|\tanh [(q_l^{-}-q_j^{+})/2]|,\ \ \ l=1,\ldots,N_{-}.
\end{gather}
Finally, we introduce real numbers
$\theta_j^{+}$, $j=1,\ldots,N_{+}$, $\theta_l^{-}$, $l=1,\ldots,N_{-}$, by writing
$\nu_n$ as
\begin{gather}
\nu_j =\frac{2V_j^{+}(q)}{\cosh (q^{+}_j)}\exp(\theta_j^{+}),\qquad
j=1,\ldots,N_{+},
\\
\nu_{N_{+}+l} =\frac{2V_l^{-}(q)}{\cosh (q^{-}_l)}\exp(\theta_l^{-}),\qquad
l=1,\ldots,N_{-}.
\end{gather}

With this reparametrization of the spectral data in effect, the
tau-function $\tau(x)$ depends on points in the
phase space
\begin{gather}
\Omega \equiv   \Big\{ (q_1^{+},\ldots,
q_{N_{-}}^{-},\theta^{+}_1,\ldots,\theta_{N_{-}}^{-})\in{\mathbb R}^{2N} \mid
q_{N_{+}}^{+}<\cdots <q_1^{+}, \nonumber \\
\phantom{\Omega     \equiv}  q_{N_{-}}^{-}<\cdots <q_1^{-},
q_j^{+}\ne q_l^{-},\ j=1,\ldots,N_{+},\ l=1,\ldots,N_{-}\Big\}\label{Omega}
\end{gather}
of the $\tilde{{\rm II}}_{{\rm rel}}(\tau
=\pi/2)$ system studied in Ref.~\cite{aa2}. The crux is now that it can be
rewritten as
\begin{equation}\label{tauL}
\tau(x)=|{\bf 1}_N+L(x+i/2)|,
\end{equation}
where
\begin{equation}
L(x)\equiv
{\mathcal L}(q^{+},q^{-},\theta_1^{+}-2\alpha_1^{+}x,\ldots,\theta_{N_{-}}^{-}-2\alpha_{N
_{-}}^{-}x),
\end{equation}
and ${\mathcal L}$ may be viewed as the Lax matrix  of this integrable $N$-particle
system. Specifically, we have
(cf.~II~Section~5)
\begin{equation}\label{cL}
{\mathcal L}(q,\theta)\equiv
{\mathcal C}(q^{+},q^{-}){\mathcal D}(q^{+},q^{-},\theta^{+},\theta^{-}),
\end{equation}
where ${\mathcal D}$ is the diagonal matrix
\begin{gather}
{\mathcal D}\equiv{\rm diag}\Big(\exp(\theta_1^{+})V_1^{+}(q),\ldots,
\exp(\theta_{N_{+}}^{+})V_{N_{+}}^{+}(q),\nonumber\\
\phantom{{\mathcal D}\equiv} \exp(\theta_1^{-})V_1^{-}(q),\ldots,
\exp(\theta_{N_{-}}^{-})V_{N_{-}}^{-}(q)\Big),\label{cD}
\end{gather}
and ${\mathcal C}$ is the Cauchy matrix
\begin{gather}\label{cC1}
{\mathcal C}_{jk}\equiv 1/\cosh [(q_j^{+}-q_k^{+})/2],
\\
\label{cC2}
{\mathcal C}_{N_{+}+l,N_{+}+m}\equiv 1/\cosh [(q_l^{-}-q_m^{-})/2],
\\
\label{cC3}
{\mathcal C}_{N_{+}+l,k}\equiv -i/\sinh [(q_l^{-}-q_k^{+})/2],
\\
\label{cC4}
{\mathcal C}_{j,N_{+}+m}\equiv i/\sinh [(q_j^{+}-q_m^{-})/2],
\end{gather}
with $j,k=1,\ldots,N_{+}$, $l,m=1,\ldots,N_{-}$.
We are now prepared to prove the lemmas.

\medskip

\noindent
{\bf Proofs of Lemmas \ref{lemma:A.1}--\ref{lemma:A.2}.} We fix $\xi\in{\mathbb R}$ and note that we may write
\begin{equation}
\tau(\xi -i/2+i\eta)= |{\bf 1}_N+L(\xi)U^{*}(\eta)|,
\end{equation}
where $U(\eta)$ is the diagonal unitary matrix
\begin{equation}
U(\eta)\equiv {\rm diag}
(\exp(2i\alpha_1^{+}\eta),\ldots,\exp(2i\alpha_{N_{-}}^{-}\eta)),\qquad
\eta\in{\mathbb R}.
\end{equation}
Putting
\begin{equation}
F(\eta)\equiv |U(\eta)+L(\xi)|,
\end{equation}
we should show that $F(\eta)$ does not vanish for any $\eta\in [-1/2,1/2]$.
For this purpose we first note that
\begin{equation}
{\mathcal D}(\xi)\equiv {\rm diag}
(\exp(\theta^{+}_1-2\alpha_1^{+}\xi)V_1^{+},\ldots,
\exp(\theta_{N_{-}}^{-}-2\alpha_{N_{-}}^{-}\xi)V_{N_{-}}^{-})
\end{equation}
is positive. Similarity transforming $L(\xi)$ with the positive diagonal
matrix ${\mathcal D}(\xi)^{1/2}$, we obtain the
symmetrized Lax matrix
\begin{equation}
L^s(\xi)\equiv {\mathcal D}(\xi)^{1/2}{\mathcal C}{\mathcal D}(\xi)^{1/2}.
\end{equation}
Since $U(\eta)$ is diagonal, we have
\begin{equation}\label{Fl}
F(\eta)=|U(\eta)+L^s(\xi)|.
\end{equation}

Consider now the inner product $(f,L^s(\xi)f)$ for a nonzero $f\in{\mathbb C}^N$.
When $N_{+}$ or $N_{-}$ vanishes, the
matrix ${\mathcal C}$ is positive, so that $(f,L^s(\xi)f)>0$. With (\ref{albig}) in
force, we also have the inequalities
\begin{equation}
\delta \, {\rm Im}\,  (f,U(\eta)f)>0,\qquad \delta \eta\in (0,1/2],\qquad f\ne 0,\qquad \delta
=+,-.
\end{equation}
Therefore, ${\rm Im}\,  (f,(U(\eta)+L^s(\xi))f)$ does not vanish for
$|\eta|\in(0,1/2]$ and $f\ne 0$. This entails that
$U(\eta)+L^s(\xi)$ is regular for $|\eta|\in(0,1/2]$. Since ${\bf
1}_N+L^s(\xi)$ is also regular, $F(\eta)$
(\ref{Fl}) does not vanish for $\eta\in[-1/2,1/2]$, and so Lemma~\ref{lemma:A.1} follows.

The restrictions (\ref{ass2}) in Lemma~\ref{lemma:A.2} amount to
\begin{equation}
\alpha_n^{\delta}\in(0,\pi/2],\qquad
 n=1,\ldots,N_{\delta},\qquad \delta =+,-.
\end{equation}
Hence we deduce
\begin{equation}\label{repo}
{\rm Re} (f,U(\eta)f)\ge 0,\qquad |\eta|\le 1/2,\qquad f\in{\mathbb C}^N.
\end{equation}
Now for $N_{+}N_{-}>0$ it is no longer true that ${\mathcal C}$ is positive. But in
view of (\ref{cC1})--(\ref{cC4}) we
may write
\begin{equation}
{\mathcal C}={\mathcal C}_{++}+{\mathcal C}_{--}+i{\mathcal C}^s,
\end{equation}
with
${\mathcal C}_{++}+{\mathcal C}_{--}$ positive and
 ${\mathcal C}^s$ real and symmetric. From this it
readily follows that we have
\begin{equation}
{\rm Re}\, (f,L^s(\xi)f)>0,\qquad f\ne 0.
\end{equation}
Combined with (\ref{repo}), this entails that $U(\eta)+L^s(\xi)$ is regular
for $|\eta|\le 1/2$. Hence
Lemma~\ref{lemma:A.2} follows.\hfill \qed

\medskip

We would like to add that this proof involves only superficial features of
the symmetrized Lax matrix.
Presumably, the far more detailed information obtained in Ref.~\cite{aa2}
can be used to relax the requirement~(\ref{ass2}).

\subsection*{Acknowledgments}

This paper was completed during our stay at the Newton Institute for
Mathematical Sciences (September--October
2001, Integrable Systems Programme). We would like to thank the Institute
for its hospitality and financial
support.

\label{Ruijsenaars-lastpage}


\begin{thebibliography}{99}
\small

\bibitem{r=0I} Ruijsenaars S~N~M, Reflectionless Analytic
Difference Operators I. Algebraic Framework,  {\it J.~Nonlin.
Math. Phys.}  {\bf 8} (2001), 106--138.

\bibitem{r=0II} Ruijsenaars~S~N~M, Reflectionless Analytic
Difference Operators II. Relations to Soliton Systems, {\it J. Nonlin.
Math. Phys.}  {\bf 8} (2001), 256--287.

\bibitem{hilb} Ruijsenaars S~N~M, Hilbert Space Theory
for Reflectionless Relativistic Potentials,
{\it Publ. RIMS Kyoto Univ.} {\bf 36} (2000), 707--753.

\bibitem{NE00} Ruijsenaars~S~N~M, Self-Adjoint A$\Delta$Os with Vanishing
Reflection, {\it Theor. Math. Phys.} {\bf 128} (2001), 933--945.

\bibitem{rs3}Reed M and Simon B, Methods of
Modern Mathematical Physics. III. Scattering Theory,
Academic Press, New York, 1979.


\bibitem{rs2} Reed M and Simon B, Methods of
Modern Mathematical Physics. II. Fourier Analysis,
Self-Adjointness, Academic Press, New York, 1975.


\bibitem{aa2} Ruijsenaars~S~N~M, Action-Angle Maps and
Scattering Theory for Some Finite-Dimensional Integrable
Systems II. Solitons, Antisolitons, and Their Bound States,
{\it Publ. RIMS Kyoto Univ.} {\bf 30} (1994), 865--1008.

\end{thebibliography}
\end{document}